\documentclass[prd,superscriptaddress,nofootinbib]{revtex4-2}
\usepackage{color}
\usepackage{bm}
\usepackage{graphicx}  
\usepackage{dcolumn}   
\usepackage{bm}        
\usepackage[english]{babel}
\usepackage{booktabs,multirow,array}
\usepackage[bottom]{footmisc}
\usepackage{dblfnote}
\usepackage{mathtools}
\usepackage{amsmath}
\DFNalwaysdouble 
\usepackage[colorlinks]{hyperref}
\definecolor{LinkColor}{rgb}{0.75, 0, 0}
\definecolor{CiteColor}{rgb}{0, 0.5, 0.5}
\definecolor{UrlColor}{rgb}{0, 0, 0.75}
\hypersetup{linkcolor=LinkColor}
\hypersetup{citecolor=CiteColor}
\hypersetup{urlcolor=UrlColor}

\def\RN{Reissner-Nordstr\"{o}m }

\labelformat{section}{Section #1} 
\labelformat{subsection}{Section #1} 
\labelformat{subsubsection}{Section #1}
\labelformat{subsubsubsection}{Section #1}
\labelformat{equation}{Eq.~(#1)} \labelformat{figure}{Fig.~#1} 
\labelformat{subfigure}{Fig.~\thefigure#1} 
\labelformat{table}{Table~#1} 
\labelformat{appendix}{Appendix #1}
\usepackage[normalem]{ulem} 

\newcommand{\pyring}{\texttt{PyRing}}
\newcommand{\pseob}{\texttt{pSEOBNRv4HM}}
\newcommand{\imr}{\texttt{IMR}}

\begin{document}
 \title{\bf Constraining extra dimensions using observations of black hole quasi--normal modes}
\author{Akash K Mishra}
\email{akash.mishra@iitgn.ac.in}
\affiliation{Indian Institute of Technology, Gandhinagar-382355, Gujarat, India}
\author{Abhirup Ghosh}
\email{abhirup.ghosh@aei.mpg.de}
\affiliation{Max Planck Institute for Gravitational Physics (Albert Einstein Institute), Am M\"uhlenberg 1, Potsdam 14476, Germany}
\author{Sumanta Chakraborty}
\email{sumantac.physics@gmail.com}
\affiliation{School of Physical Sciences, Indian Association for the Cultivation of Science, Kolkata-700032, India}


\begin{abstract}
The presence of extra dimensions generically modify the spacetime geometry of a rotating black hole, by adding an additional hair, besides the mass $M$ and the angular momentum $J$, known as the `tidal charge' parameter, $\beta$. In a braneworld scenario with one extra spatial dimension, the extra dimension is expected to manifest itself through --- (a) negative values of $\beta$, and (b) modified gravitational perturbations. This in turn would affect the quasi-normal modes of rotating black holes. We numerically solve the perturbed gravitational field equations using the continued fractions method and determine the quasi-normal mode spectra for the braneworld black hole. We find that increasingly negative values of $\beta$ correspond to a diminishing imaginary part of the quasi-normal mode, or equivalently, an increasing damping time. Using the publicly available data of the properties of the remnant black hole in the gravitational wave signal GW150914, we check for consistency between the predicted values (for a given $\beta$) of the frequency and damping time of the least-damped $\ell=2,m=2$ quasi-normal mode and measurements of these quantities using other independent techniques. We find that it is highly unlikely for the tidal charge, $\beta \lesssim -0.05$, providing a conservative limit on the tidal charge parameter. Implications and future directions are discussed.
\end{abstract}

\maketitle
\section{Introduction}
Our present understanding of gravitational interaction is best described by Einstein's theory of general relativity (GR) \cite{Einstein:1916cc,Einstein:1918btx,Hawking:1973uf,Wald:1984rg}. The results derived from GR are in excellent agreement with observations across a large range of length scales~\cite{Will:2005va, Berti:2015itd,Berti:2018cxi,Berti:2018vdi}; from weak field tests of gravity, like perihelion precession and lensing, to the strong field, such as gravitational waves (GWs) from merging compact objects~\cite{LIGOScientific:2019fpa,Abbott:2020jks,TheLIGOScientific:2017qsa} or observations of black hole (BH) shadows \cite{Akiyama:2019cqa}. However despite its success, GR faces severe theoretical challenges and there are reasons to believe that it should (possibly) be modified in the very short and very long length scales. These challenges include the incompatibility between GR and quantum theory \cite{Birrell:1982ix}, presence of spacetime singularities \cite{PhysRevLett.14.57, PhysRevD.14.2460}, the violation of strong cosmic censorship conjecture leading to a loss of determinism \cite{Cardoso:2017soq,Rahman:2018oso} and, of course, the late time acceleration of the universe and the cosmological constant problem \cite{RevModPhys.61.1, Padmanabhan:2002ji,Carroll:2000fy}, to name a few. Though it is expected that a fully consistent quantum theory of gravity would ultimately overcome these problems, in the absence of such a theory, an effective approach is to look for possible alternatives to GR, which may address some of the issues listed above. This has led to the development of several classes of modified theories of gravity and exploring them in detail has been one of the central themes of research in gravitational physics (for a small sample of works, see \cite{Clifton:2011jh,Capozziello:2011et,Nojiri:2010wj,Gleyzes:2014dya}). 

In general, any correction term to GR, which is consistent with diffeomorphism symmetry, may contribute to the classical gravitational action. As a result, there is no unique way to modify GR. However, if diffeomorphism invariance is the only criteria to add new terms to the gravitational action, there would have been an infinite number of such modified theories of gravity and hence the task of identifying the correct Lagrangian through a finite number of observations would appear impossible. In this apparently grim situation, the Ostrogradsky instability helps to eliminate all modified theories of gravity yielding higher order field equations \cite{Woodard:2015zca}, and restricts the form of the correction terms one may add over and above the Einstein-Hilbert term in GR. Further constraints on these restricted class of theories, with second order field equations, can be derived by checking their consistency with the observations in the weak as well as in the strong field regime. In particular, since these corrections over GR are expected to be dominant in the high energy/small length scale regime, it is necessary to compare various predictions of such modified theories with some strong gravity observations, as and when they become available. The detections of GWs from coalescences of compact binary sources like neutron stars and/or BHs~\cite{LIGOScientific:2018mvr,Abbott:2020niy} by the LIGO-Virgo detectors~\cite{TheLIGOScientific:2014jea,TheVirgo:2014hva} provide an excellent opportunity to test, at an unprecedented level, predictions of GR in the highly dynamical strong--field regimes of gravity \cite{LIGOScientific:2019fpa,Abbott:2020jks,TheLIGOScientific:2017qsa}. In particular, GWs from these merger events allow us to not only test GR in regimes of extreme gravity, but also constrain parameters of alternative theories. These studies often lead to several interesting bounds on the magnitudes of possible deviation from GR (for a small sample of references, see \cite{Baker:2017hug,Jana:2017ost,Creminelli:2017sry, Nojiri:2017hai,Akrami:2018yjz,Chakravarti:2019aup,Ghosh:2019twk,Yu:2016tar, Lin:2020wnp, Wang:2021uuh} and references therein). As an aside, note that, besides GWs, the recent observation involving BH shadow is also a strong field test of gravitational interaction, which can also provide constraints on deviations from GR \cite{Psaltis:2020lvx,Zhu_2018,Banerjee:2019nnj}.  

In this work, we concentrate on the modifications of GR due to the presence of an extra spatial dimension \cite{Maartens:2003tw,Csaki:2004ay,PerezLorenzana:2005iv,Kanti:2004nr} and try to constrain the same using GW observations. Inclusion of an extra spatial dimension in our usual four dimensional spacetime has a long history, starting from the attempt of Kaluza and Klein to unify gravity and electromagnetism (for a review, see \cite{Overduin:1998pn}). Extra dimensional scenarios came to the limelight again when it was realized that these models can address the long standing gauge hierarchy problem in high energy physics. The huge gap $\sim \mathcal{O}(10^{17})$, between the electroweak scale and the Planck scale --- leading to extreme fine-tuning --- is known as the gauge hierarchy problem \cite{Antoniadis:1998ig,Randall:1999ee}. This fine-tuning is essential in order to keep the mass of the Higg's Boson in the electroweak scale and achieving consistency with the LHC results~\cite{Aad:2012tfa,Chatrchyan:2012ufa}. Presence of extra spatial dimensions, either through large volume \cite{Antoniadis:1998ig,ArkaniHamed:1998rs} or through exponential warping \cite{Randall:1999ee,Randall:1999vf}, can reduce the four dimensional Planck scale to electroweak scale and hence the fine tuning/gauge hierarchy problem can be avoided. Latter studies have shown several other contexts having interesting applications of the higher dimensional scenario, which includes --- BHs \cite{Harko:2004ui,Aliev:2005bi,Chakraborty:2015taq,Chakraborty:2014xla,Dadhich:2000am,Chamblin:1999by,Chamblin:2000ra,Emparan:1999wa,Nakas:2020sey, Nakas:2021srr}, cosmology \cite{Csaki:1999mp,Csaki:1999jh}, GWs \cite{Abbott:2018lct,Visinelli:2017bny,Chakraborty:2017qve,Chakravarti:2018vlt,Chakravarti:2019aup,Toshmatov:2016bsb,Rahman:2018oso,Dey:2020pth,deOliveira:2020lzp,Dey:2020lhq} among others. In most of these higher dimensional scenario, the effective gravitational dynamics in four dimensions, which is a hypersurface in the full higher dimensional spacetime will be different from that of Einstein gravity. The fact that we are actually living in a higher dimensional spacetime must appear somehow in our effective four dimensional gravitational dynamics. It is worth mentioning that except gravity, other fields are taken to be confined to the four dimensional spacetime, while gravity alone can probe the extra dimensions. For our purpose it will suffice to consider a five dimensional spacetime with a single extra spatial dimension, referred to as the \emph{bulk} spacetime, while our four dimensional universe is known as the \emph{brane}. It is important to emphasize that the braneworld scenario considered here is general enough to encompass the situation in which the extra spatial dimension need not be compact. For simplicity we assume Einstein gravity in the bulk spacetime, in which case the gravitational dynamics on the brane is governed by an appropriate projection of the bulk Einstein's equations on the brane, which will have corrections over and above the Einstein term. These corrections are precisely what we wish to explore. Interestingly, the effective gravitational field equations on the brane exhibits localized BH solutions, which resemble the \RN and the Kerr-Newman solutions of GR, with the crucial difference being the charge term (often referred to as \emph{tidal charge}) taking negative values \cite{Shiromizu:1999wj,Dadhich:2000am,Aliev:2005bi,Aliev:2009cg}. Note that the tidal charge parameter is sourced by the extra spatial dimension, such that in the GR limit it identically vanishes. Previous works have also reported interesting constraints on the tidal charge parameter and consequently on the extra spatial dimension \cite{Horvath:2012ru, Zakharov:2018awx, Banerjee:2019sae, Banerjee:2019nnj, Chakravarti:2019aup, Neves:2020doc,Chakraborty:2021gdf}. However as we will see none of these constraints are as robust as we will derive in the present work. In what follows, we will develop the formalism to constrain the tidal charge parameter of a rotating braneworld BH using publicly available measurements of GW observations. In particular, by using the measurements of the remnant properties and (complex) quasi-normal mode (QNM) frequencies of the ringdown signal in the first-ever gravitational wave event GW150914 \cite{Abbott:2016blz,TheLIGOScientific:2016src}, we obtain a novel upper bound on the magnitude of the tidal charge.

The rest of the article is arranged as follows: In \ref{sec:brane} we briefly review the effective field equations on the brane and the associated rotating BH solution. The computation of the QNMs associated with a rotating braneworld BH, using the continued fractions method has been presented in \ref{sec:method}. Finally the comparison with the GW150914 event and the resulting constraint has been presented in \ref{sec:bounds}. We conclude with a discussion on our results and possible future directions. 

\emph{Notations and Conventions:} In this work we will follow the mostly positive signature convention, i.e., the flat spacetime Minkowski metric in four dimensions takes the form, $\textrm{diag}(-1,1,1,1)$. Indices referring to higher dimensional spacetime are denoted by uppercase Roman letters and the indices for the four dimensional spacetime are represented by Greek letters. We also set the fundamental constants to unity, i.e., $c=1=G$.

\section{Brief review of rotating braneworld Black hole}\label{sec:brane}

In this section we will briefly review the effective gravitational field equations on the four dimensional brane and the geometry of rotating BH solutions arising from the field equations. As emphasized earlier, we consider the gravitational interaction in the five dimensional bulk spacetime to be described by Einstein gravity. However, the effective four dimensional description of the gravitational interaction will not be governed by Einstein's equations, rather there will be corrections over and above the same. These corrections arise as we project the five dimensional Einstein's equations on the four dimensional brane hypersurface using an appropriate projector $h^{A}_{B}=\delta^{A}_{B}-n^{A}n_{B}$, where $n_{A}$ is the unit normal to the brane hypersurface, satisfying $n_{A}n^{A}=1$. The projection of the five dimensional Einstein tensor $G_{AB}$ on the four dimensional brane uses the Gauss-Codazzi and the Mainardi relations, connecting geometrical quantities in the full spacetime to geometrical quantities in a lower dimensional hypersurface. This results into the following effective gravitational field equations on the brane \cite{Shiromizu:1999wj},
\begin{align}
~^{(4)}G_{\mu \nu}+E_{\mu \nu}=8\pi G T_{\mu \nu}+\Pi_{\mu \nu}~.
\end{align}
Here, $E_{\mu \nu}=W_{ABCD}n^{A}e^{B}_{\mu}n^{C}e^{D}_{\nu}$ is the electric part of the bulk Weyl tensor $W_{ABCD}$, with $T_{\mu \nu}$ being the matter energy-momentum tensor on the brane. Additionally, the tensor $\Pi_{\mu \nu}$ appearing in the effective gravitational field equations presented above, is a quadratic combination of $T_{\mu \nu}$, e.g., it involves terms like, $T_{\mu \alpha}T^{\alpha}_{\nu}$, $TT_{\mu \nu}$ etc. Since we will be interested in vacuum four dimensional spacetime, the matter energy-momentum tensor on the brane would vanish identically and hence the $\Pi_{\mu \nu}$ term will not contribute in the present context. Thus for vacuum brane, the gravitational dynamics is governed by the following effective equations,
\begin{align}\label{eff_eq}
~^{(4)}G_{\mu \nu}+E_{\mu \nu}=0~.
\end{align}
Thus for our purpose the bulk Weyl tensor plays the most important role and is the factor responsible for modifications to the Einstein's equations. Note that due to symmetry properties of the Weyl tensor, $E_{\mu \nu}$ is traceless and due to Bianchi identity it is also divergence free. Both of these properties hold true for electromagnetic stress-tensor as well and hence the BH solutions arising out of the above effective gravitational field equations very much resemble the Kerr-Newman family of BHs. With one crucial sign difference --- the electromagnetic stress-energy tensor appears on the right hand side of the field equations --- while here $E_{\mu \nu}$ appears on the left hand side, as evident from \ref{eff_eq}. In particular, the rotating BH solution arising out of the effective field equations on the brane takes the following form \cite{Dadhich:2000am,Aliev:2005bi,Aliev:2009cg},  
\begin{align}
ds^2 = -\frac{\Delta}{\Sigma}(dt- a \sin^2\theta\, d\phi)^2+\,\Sigma\left[\frac{dr^2}{\Delta}+\,d\theta^2\right] + \frac{\sin^2\theta}{\Sigma}\left[a\,dt-(r^2+a^2)d\phi\right]^2~,
\end{align}
where, $a$ and $M$ are the spin and mass of the BH respectively, and $\Delta\equiv r^{2}+a^{2}-2Mr+q$ and $\Sigma\equiv r^{2}+a^{2}\cos^{2}\theta$. Note that, for the case of Kerr-Newman BH, the parameter $q$ can be identified with the square of the BH charge, i.e., $q|_{\rm KN}=Q^{2}$. However, in the braneworld scenario, $q$ represents the tidal charge parameter and hence it can take negative values as well. This is the key feature for the braneworld BHs, which we wish to explore in detail in this work from the perspective of QNMs. 

In addition, we briefly discuss about some other interesting properties of this solution. Since the horizons of the above solution are located at, $r_{\pm}=M\pm\sqrt{M^{2}-a^{2}-q}$, in the non-rotating case with $q<0$, there is only \emph{one} horizon, in sharp contrast with the case of Reissner-Nordstr\"{o}m BH. Similarly, in the rotating case, for the existence of horizons, the rotation parameter must be bounded by $(a/M)^{2}\leq 1-(q/M^{2})$, which can be larger than unity for negative values of $q$. This is again in striking contrast to the case of a Kerr-Newman BH, for which the value of the dimensionless rotation parameter $(a/M)$ is strictly less than unity. Furthermore, negative value of the tidal charge has implications in various other astrophysical scenarios, e.g., --- (a) the tidal love number of a braneworld BH is \emph{non-zero} \cite{Chakravarti:2018vlt}, (b) braneworld BHs cast a bigger shadow and is consistent with the shadow measurement of the supermassive BH M87* \cite{Banerjee:2019nnj}, (c) continuum spectrum as well as quasi-periodic oscillations from accretion disks favours the presence of extra dimensions \cite{Banerjee:2019sae,Banerjee:2021aln}. Motivated by these results, we concentrate, in this work, on the implications of a negative tidal charge parameter on the QNMs. We present the computation of the QNMs for rotating braneworld BH in the next section.    

\section{Quasi-normal modes of a rotating braneworld black hole}\label{sec:method}

The spacetime metric of a rotating BH on the vacuum brane embedded in a higher dimensional spacetime, along with its physical characteristics have been elaborated in the previous section. In this section, we will outline the method for the determination of the BH QNMs. Unlike the previous section, here we will assume $2M=1$ for simplicity of the analysis, however all the factors involving the mass of the BH will be restored, while comparing with the GW observations in the next section. 

We focus on the case of linear gravitational perturbations in the background of a rotating braneworld BH, which, unlike the case of a Kerr BH \cite{Teukolsky:1973ha}, is generically non-separable \cite{Berti:2005eb}. In the context of Kerr-Newman BH, the separability is achieved under the Dudley-Finnley approximation \cite{Dudley:1978vd, PhysRevLett.38.1505}, where the electromagnetic charge was assumed to be small. However, this approximation is not a good one, as demonstrated in \cite{Mark:2014aja} and further corroborated by the results of \cite{Dias:2015wqa}. However, in the present context, we assume that the gravitational perturbations on the brane, keep the contribution from the bulk geometry, i.e., $E_{\mu \nu}$ unchanged. This is because, following \cite{Kanno:2003au, Kanno:2003sc} one can argue that the perturbation of the bulk Weyl tensor has the form $\delta E_{\mu \nu}=(\ell/L)E_{\mu \nu}$, where $\ell$ is a characteristic length scale of the bulk geometry, while $L$ is a characteristic length scale of the black hole on the brane. It is obvious that $(\ell/L)\ll 1$. Thus we are assuming that the bulk curvature scale is much larger than the curvature on the brane, which is a much more robust approximation than that of small electromagnetic charge due to Dudley and Finnley.  Under this assumption, for reasonable values of the tidal charge parameter $q$, the radial and the angular part of the gravitational perturbation also separates, identical but very different in spirit to the Kerr-Newman spacetime \cite{Berti:2005eb}. This allows us to determine the QNMs of the background BH spacetime under perturbations. It is worth mentioning that this also ensures the separability of generic spin `s' perturbation. 

Given the separability of a generic spin `s' perturbation $\Psi_{s}(t,r,\theta,\phi)$, it follows that the perturbation can be decomposed into temporal, radial, angular and azimuthal part as,
\begin{align}
\Psi_{s}(t,r,\theta,\phi)=\sum_{\ell,m}e^{-i\omega t}R_{\ell m}(r)S_{\ell m}(\theta)e^{im\phi}~,
\end{align}
where, $\ell$ is the angular momentum and $m$ is its z-component, such that $m\in (-\ell,-\ell+1,\cdots,\ell-1,\ell)$, and $\omega$ is the QNM frequency. Substituting the above spin `s' perturbation into the linearized gravitational field equations, the separated radial perturbation $R_{\ell m}(r)$ and the angular perturbation $S_{\ell m}(\theta)$ satisfies the following equations on the braneworld BH spacetime,
\begin{align}\label{angular}
\frac{d}{du}\left[\left(1-u^{2}\right)\frac{dS_{\ell m}}{du}\right] + \left[(a\omega u)^2 - 2 a \omega s u +s +A_{\ell m} - \frac{(m+s u)^2}{1-u^2}\right]S_{\ell m}=0~,
\end{align}
\begin{align}\label{radial}
\Delta\left(\frac{d^2R_{\ell m}}{dr^2}\right)+(s+1)(2r-1)&\Big(\frac{dR_{\ell m}}{dr}\Big)
+\biggl[-\left\{a^2+q+\left(r-1\right) r\right\}\left\{A_{\ell m}+\omega \left(a^2\omega-2am-4irs\right)\right\}
\nonumber
\\&-i\left(2r-1\right)s\left\{\omega\left(a^2+r^2\right)-am\right\}+\left\{am-\omega \left(a^2+r^2\right)\right\}^2\biggl]R_{\ell m}=0~.
\end{align}
Here the spin parameter $s$ takes values $(0,-1,-2)$ for scalar, electromagnetic and gravitational perturbations respectively and $u \equiv \cos\theta$. The separation constant $A_{\ell m}$ appearing in both the radial and angular equation reduces to $\ell(\ell+1)-s(s+1)$ in the limit of vanishing rotation parameter $a$. The above pair of differential equations can be solved to obtain $(\omega, A_{\ell m})$ by setting appropriate regularity and boundary conditions.  

The relevant boundary condition for the angular equation is the finite behaviour of $S_{\ell m}$ at the regular singular points of the angular equation presented in \ref{angular}, which are located at $(u=1,-1)$. Therefore we will employ the Leaver's method \cite{Leaver:1985ax} for solving these differential equations, which effectively is equivalent to finding a series solution to the angular differential equation, given by \ref{angular}. Given the regular singular points, the series solution to the angular equation can be expressed as, 
\begin{align}\label{ang_frob_series}
S_{\ell m}(u)=e^{a\omega u}(1+u)^{k_1}(1-u)^{k_2}\sum_{n=0}^\infty c_n (1+u)^n~,
\end{align}
where, $k_1=\frac{1}{2}|m-s|$ and $k_2=\frac{1}{2}|m+s|$. The expansion coefficients $c_n$, appearing in the above series solution, are related to each other by a three term recurrence relation, which takes the following form,
\begin{align}
\alpha_{n}^{(\theta)}c_{n+1}+\gamma_{n}^{(\theta)}c_{n}+\delta_{n}^{(\theta)}c_{n-1}=0~, \qquad (n=1,2,3,\ldots)~.
\end{align}
The coefficients $\alpha_{n}^{(\theta)}$, $\gamma_{n}^{(\theta)}$ and $\delta_{n}^{(\theta)}$, appearing in the above recurrence relation for the angular equation are of the following form,
\begin{align}
\alpha_{n}^{(\theta)}&=-2(n+1)(2k_1+n+1)~,
\\
\gamma_{n}^{(\theta)}&=-\left[a^2\omega^2+\left(s+1\right)s+A_{\ell m}\right]+2n\left(-2a\omega+k_1+k_2+1\right)
\nonumber
\\
&\hskip 2 cm -\left[2a\omega\left(2k_1+s+1\right)-\left(k_1+k_1\right)\left(k_1+k_1+1\right)\right]+(n-1)n~,
\\
\delta_{n}^{(\theta)}&=2a\omega\left(k_1+k_2+n+s\right)~.
\end{align}
It is to be noted that the above expressions are identical to those in \cite{Leaver:1985ax}.
Alike the series solution to the angular equation, one can obtain a series solution to the radial equation by setting similar boundary conditions --- (a) perturbations are purely ingoing at the BH horizon and (b) perturbations are purely outgoing at infinity. Thus the series solution, with regular singular points at $r=r_{\pm}$, takes the following form,
\begin{align}
R_{\ell m}(r)=e^{i\omega r}\left(r-r_{+}\right)^{-s-i\sigma_+}\left(r-r_{-}\right)^{-1-s+i\omega+i\sigma_{+}}\sum_{n=0}^\infty d_{n} \left(\frac{r-r_+}{r-r_{-}}\right)^{n}~,
 \end{align}
where, $r_{\pm}=(1/2)(1\pm b)$ are the horizon locations. Here, $b\equiv \sqrt{1-4(a^2+q)}$ and $\sigma_{+} \equiv (1/b)[\omega(r_{+}-q)-am]$. The coefficients $d_{n}$ also satisfies a three term recurrence relation, which can be obtained by substituting the above series solution for the radial perturbation $R_{\ell m}(r)$ in the radial perturbation equation, given by \ref{radial}, which yields,
\begin{align}
\alpha_{n}^{(r)} d_{n+1}+\gamma_{n}^{(r)}d_{n}+\delta_{n}^{(r)}d_{n-1}=0~, \qquad (n=1,2,3,\ldots)~.
\end{align}
with the coefficients $\alpha_{n}^{(r)}$, $\gamma_{n}^{(r)}$ and $\delta_{n}^{(r)}$ are given by,
\begin{align}
\alpha_{n}^{(r)}&=(n+1)\Big[-2i q\omega \sqrt{-4a^2-4q+1}+i\omega\sqrt{-4a^2-4q +1}-2iam\sqrt{-4a^2-4q+1}
\nonumber
\\
&\hskip 2 cm +(n+1) \left(4 a^2+4q -1\right)\Big]-(n+1)\left(4 a^2+4q -1\right)\left(s+i\omega\right)~,
\\
\gamma_{n}^{(r)}&=-4a^4\omega^2-8 a^3m\omega-4\omega^2\sqrt{-4a^2-4q+1}+12q \omega^2\sqrt{-4a^2-4q+1}
\nonumber
\\
&\hskip 2 cm -2i\omega\sqrt{-4a^2-4q +1}+6iq\omega\sqrt{-4a^2-4q+1}-4in \omega \sqrt{-4a^2-4q+1}
\nonumber
\\
&\hskip 2 cm +2am\left[i\sqrt{-4a^2-4q+1}+2\sqrt{-4 a^2-4q +1}\left(\omega +i n\right)-4q\omega+\omega \right]
\nonumber
\\
&\hskip 2 cm +12iq n\omega \sqrt{-4a^2-4q +1}+\left(1-4q\right)\left[A_{\ell m}+2n\left(n+1\right)+s+1\right]
\nonumber
\\
&\hskip 2 cm -4\left(4q^2-5q+1\right)\omega^2+2i (4q -1) (2n+1) \omega
\nonumber
\\
&\hskip 2 cm +a^{2}\Big[\omega\left\{\omega \left(8 \sqrt{-4 a^2-4q+1}-20q+17\right)+4 i \left(\sqrt{-4 a^2-4q +1}+2\right)\right\}
\nonumber
\\
&\hskip 2.5 cm +8in\left\{\omega \left(\sqrt{-4 a^2-4q+1}+2\right)+i\right\}\Big]-\left(4A_{\ell m}+8 n^2+4 s+4 \right)a^2~,
\\
\delta_{n}^{(r)}&=\left(n-2i\omega\right)\Big[-2iq\omega\sqrt{-4a^2-4q+1}+i\omega\sqrt{-4a^2-4q+1}-2iam \sqrt{-4a^2-4q+1}
\nonumber
\\
&\hskip 2 cm +4a^2\left(n+s-i\omega\right)\Big]+\left(n-2i\omega\right)\left(4q-1\right)\left(n+s-i\omega\right)~.
\end{align}

\begin{table}[ht]
\begin{center}
\begin{tabular}{p{0.1\textwidth}>{\centering}p{0.1\textwidth}>{\centering}p{0.1\textwidth}>{\centering}p{0.1\textwidth}>{\centering}p{0.1\textwidth}>{\centering}p{0.1\textwidth}>{\centering}p{0.2\textwidth}>{\centering}p{0.15\textwidth}}
$\ell,m$ & $-\beta$ & $\omega_{\rm r}$ & -$\omega_{\rm i}$ & Damping time (ms)&  Frequency (Hz) \tabularnewline
\hline 

\multirow{16}{*}{$\ell = 2, m=2$} & \multirow{1}{*}{$0$} & \multirow{1}{*}{$1.039711$}  & $0.163690$ & $4.069 $ &$248.395 $ \tabularnewline

 &  \multirow{1}{*}{$0.05$} &  \multirow{1}{*}{$0.963809$} & $0.163932$ &  $4.063$ &$ 230.261$ \tabularnewline
 &   \multirow{1}{*}{$0.1$}  & \multirow{1}{*}{$0.905704$} & $0.162635$ &  $4.096 $ &$216.380 $\tabularnewline
 & \multirow{1}{*}{$0.15$}  & \multirow{1}{*}{$0.858954$} & $0.160679$ &  $4.146 $ &$205.211 $\tabularnewline
  & \multirow{1}{*}{$0.2$}  & \multirow{1}{*}{$0.820065$} & $0.158443$ &  $4.204 $ &$195.920 $\tabularnewline
   & \multirow{1}{*}{$0.25$}  & \multirow{1}{*}{$0.786925$} & $0.156107$ &  $4.267 $ &$188.002 $\tabularnewline
    & \multirow{1}{*}{$0.3$}  & \multirow{1}{*}{$0.758165$} & $0.153765$ &  $4.332 $ &$181.131 $\tabularnewline
     & \multirow{1}{*}{$0.35$}  & \multirow{1}{*}{$0.732845$} & $0.151463$ &  $4.398 $ &$175.082 $\tabularnewline
      & \multirow{1}{*}{$0.4$}  & \multirow{1}{*}{$0.710292$} & $0.149226$ &  $4.464 $ &$169.694 $\tabularnewline
           & \multirow{1}{*}{$0.45$}  & \multirow{1}{*}{$0.690011$} & $0.147066$ &  $4.529 $ &$164.849 $\tabularnewline
       & \multirow{1}{*}{$0.5$}  & \multirow{1}{*}{$0.671623$} & $0.144988$ &  $4.594 $ &$160.456 $\tabularnewline
       & \multirow{1}{*}{$0.55$}  & \multirow{1}{*}{$0.654837$} & $0.142992$ &  $4.658 $ &$156.446 $\tabularnewline
       & \multirow{1}{*}{$0.6$}  & \multirow{1}{*}{$0.639422$} & $0.141076$ &  $4.722 $ &$152.763 $\tabularnewline
       & \multirow{1}{*}{$0.65$}  & \multirow{1}{*}{$0.625192$} & $0.139238$ &  $4.784 $ &$149.363 $\tabularnewline
       & \multirow{1}{*}{$0.7$}  & \multirow{1}{*}{$0.611995$} & $0.137475$ &  $4.845 $ &$146.210 $\tabularnewline
       & \multirow{1}{*}{$0.75$}  & \multirow{1}{*}{$0.599706$} & $0.135782$ &  $4.906 $ &$143.274 $\tabularnewline
       & \multirow{1}{*}{$0.8$}  & \multirow{1}{*}{$0.58822$} & $0.134157$ &  $4.965 $ &$140.530 $\tabularnewline

\hline 
\multirow{16}{*}{$\ell = 3, m=3$} & \multirow{1}{*}{$0$} & \multirow{1}{*}{$1.64959$}  & $0.168181$ & $3.92107 $ &$394.109 $ \tabularnewline

 &  \multirow{1}{*}{$0.05$} &  \multirow{1}{*}{$1.54121$} & $0.16998$ &  $3.91915$ &$348.185$ \tabularnewline
 &   \multirow{1}{*}{$0.1$}  & \multirow{1}{*}{$1.45737$} & $0.169987$ &  $3.91899 $ &$348.185 $\tabularnewline
 & \multirow{1}{*}{$0.15$}  & \multirow{1}{*}{$1.38932$} & $0.169127$ &  $3.93892 $ &$331.927 $\tabularnewline
  & \multirow{1}{*}{$0.2$}  & \multirow{1}{*}{$1.33226$} & $0.167821$ &  $3.96959 $ &$318.295 $\tabularnewline
   & \multirow{1}{*}{$0.25$}  & \multirow{1}{*}{$1.283312$} & $0.166279$ &  $4.00638 $ &$306.003 $\tabularnewline
    & \multirow{1}{*}{$0.3$}  & \multirow{1}{*}{$1.240552$} & $0.164621$ &  $4.04673 $ &$296.384 $\tabularnewline
     & \multirow{1}{*}{$0.35$}  & \multirow{1}{*}{$1.20269$} & $0.162912$ &  $4.08919 $ &$287.339 $\tabularnewline
      & \multirow{1}{*}{$0.4$}  & \multirow{1}{*}{$1.168812$} & $0.161192$ &  $4.13282 $ &$279.244 $\tabularnewline
           & \multirow{1}{*}{$0.45$}  & \multirow{1}{*}{$1.138176$} & $0.159484$ &  $4.17708 $ &$271.924 $\tabularnewline
       & \multirow{1}{*}{$0.5$}  & \multirow{1}{*}{$1.11029$} & $0.157804$ &  $4.22155 $ &$265.263 $\tabularnewline
       & \multirow{1}{*}{$0.55$}  & \multirow{1}{*}{$1.08473$} & $0.156166$ &  $4.26599 $ &$259.071 $\tabularnewline
       & \multirow{1}{*}{$0.6$}  & \multirow{1}{*}{$1.061172$} & $0.154556$ &  $4.31027 $ &$253.528 $\tabularnewline
       & \multirow{1}{*}{$0.65$}  & \multirow{1}{*}{$1.035172$} & $0.152688$ &  $4.36321 $ &$247.316 $\tabularnewline
       & \multirow{1}{*}{$0.7$}  & \multirow{1}{*}{$1.01904$} & $0.151478$ &  $4.39785 $ &$243.462 $\tabularnewline
       & \multirow{1}{*}{$0.75$}  & \multirow{1}{*}{$1.000172$} & $0.150121$ &  $4.43941$ &$238.930 $\tabularnewline
       & \multirow{1}{*}{$0.8$}  & \multirow{1}{*}{$0.982296$} & $0.148577$ &  $4.48372 $ &$234.684 $\tabularnewline

\hline 
\end{tabular}
\caption{Numerical values of the real and imaginary parts of the excited QNM $(n=0,\ell=3=m)$ frequencies, along with the oscillation frequency and damping time, corresponding to the gravitational perturbation $(s=-2)$, for various values of tidal charge parameter $\beta$ have been presented, for a BH of mass $M=62 M_\odot$ and dimensionless spin parameter $\chi=0.67$. The real and imaginary parts of the QNM frequencies in natural units have been converted to oscillation frequency in Hz and damping time in ms through the following relations: $f_{n\ell m}~(\textrm{Hz})=(1/2\pi)(c^{3}/2GM)(1+z)^{-1}(\textrm{Re}~\omega_{n\ell m})$ and $\tau_{n\ell m}~(\textrm{ms})=10^{3}(2GM/c^{3})(1+z)(\textrm{Im}~\omega_{n\ell m})^{-1}$.}
\label{table-QNM-Kerr-tidal3}
\end{center}
\end{table}

\begin{figure}
\centering
\includegraphics[scale=0.5]{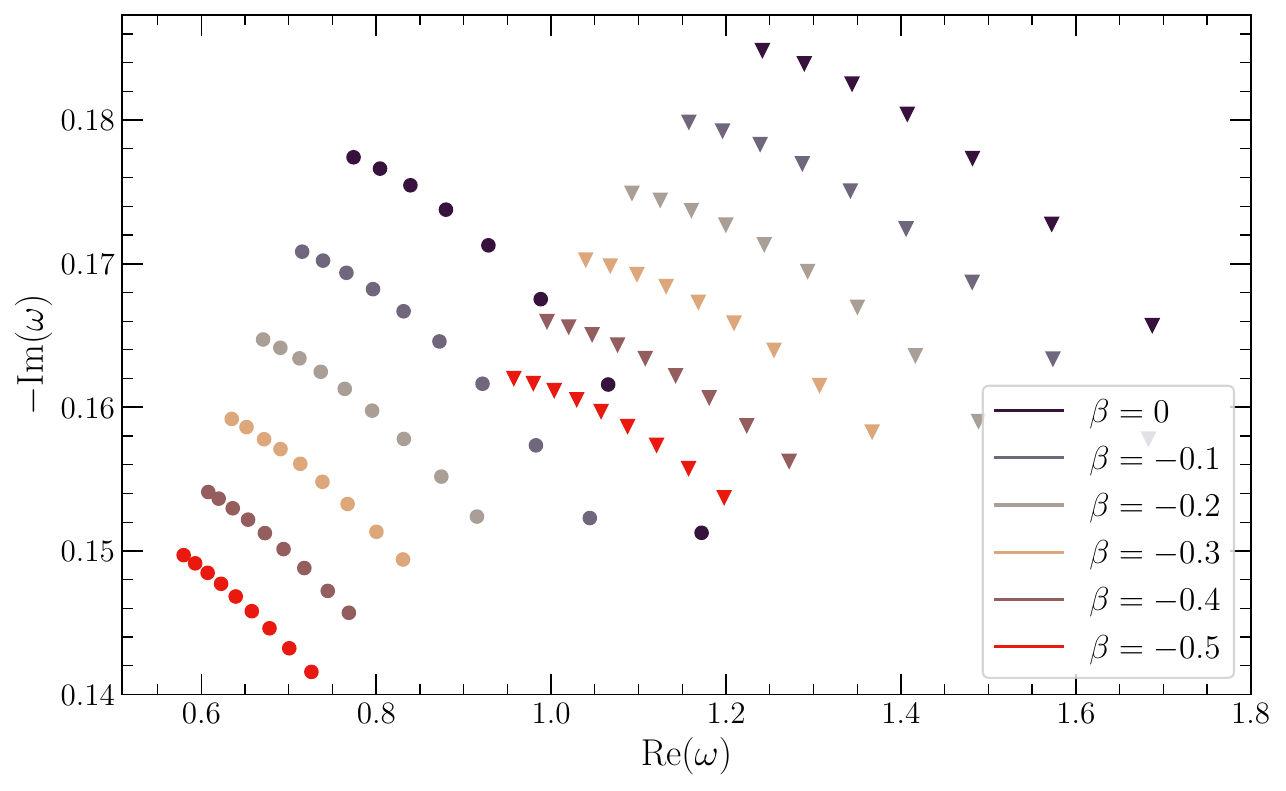}
\caption{In this figure we have plotted the real and imaginary parts of the QNM frequency $\omega_{n\ell m}$ for different choices of the tidal charge $\beta$. The circled points are for the fundamental $\ell=2=m$ mode, while the triangle-like points are for the excited $\ell=3=m$ mode. For each value of $\beta$, the points refer to the spin parameter having values, $\chi= 0.1,0.2,0.3,\cdots 0.9$ (left to right). As evident, with the increase of $|\beta|$, the decrease in $(\textrm{Re}~\omega_{n\ell m})$ is smaller than the decrease in $(\textrm{Im}~\omega_{n\ell m})$. See text for more discussions.}
\label{real_im_freq}
\end{figure}

Having derived the recurrence relations for the angular and the radial perturbation equations, let us now proceed to (numerically) solve simultaneously these three--term recurrence relations using the continued fraction method, and obtain the (complex) QNM frequencies, $\omega_{n\ell m}:=2\pi f_{n\ell m}-i\tau_{n\ell m}^{-1}$, where ($f_{n\ell m}, \tau_{n\ell m}$) represent the frequency and damping time of the ${n\ell m}$-th QNM respectively. Note that each QNM frequency is characterized by the overtone number $n$, the angular momentum $\ell$ and its z-component $m$. It is worth mentioning that the $n$ in the QNM frequency refers to the QNM overtone; not to be confused with the dummy variable used in the series expansions of the perturbations, appearing previously in this section. 

The recurrence relations for the angular and radial perturbation, depends on the mass $M$, the spin $a$ and the tidal charge parameter $q$. Hence the QNM frequencies also depend on these hairs. However it is convenient to introduce the dimensionless parameters $\chi\equiv (a/M)$ and $\beta\equiv (q/4M^{2})$ and hence the real and imaginary parts of the QNM frequencies can be expressed as,
\begin{align}
f_{n\ell m}&=f_{n\ell m}(M,\chi,\beta)~,
\label{freq_rel}
\\
\tau_{n\ell m}&=\tau_{n\ell m}(M,\chi,\beta)~. 
\label{damp_rel}
\end{align}
Therefore, given the mass $M$ and spin $\chi$ of BH, perhaps the remnant from the merger of two BHs, one can predict the oscillation frequency and damping time for different values of the tidal charge $\beta$. We calculate the predictions of the frequencies and damping times for the least-damped ($n=0$) $\ell=2,m=2$ and $\ell=3,m=3$ QNMs of a BH of mass $M=62 M_\odot$ and spin $\chi=0.67$ in \ref{table-QNM-Kerr-tidal3}. These values for the mass and spin are chosen to be close to the median values of these quantities for the remnant BH of GW150914~\cite{Abbott:2016blz}, the first GW signal observed from the merger of two (non-spinning) BHs of $\sim 30M_{\odot}$ each. We perform a more detailed consistency check between the QNM predictions and their observed estimates for GW150914 in the next section.

We also plot the real and imaginary parts of the QNM frequencies for different values of the tidal charge $\beta$ and spin $\chi$ for the fundamental $\ell=2=m$ case and the excited $\ell=3=m$ case in \ref{real_im_freq}. As evident, for a given $\beta$ with an increase of the spin $\chi$, the imaginary part of the QNM frequency decreases much slowly compared to the real part. On the other hand, for a fixed $\chi$, with an increase in the tidal charge $|\beta|$, both the imaginary and the real part decreases, but the decrease in the real part is smaller compared to the decrease in the imaginary part. As a consequence the change in the damping time $\tau_{n\ell m}$ is much smaller than in the oscillation frequency $f_{n\ell m}$. This behaviour of the oscillation frequency and damping time will be the key to constrain the tidal charge parameter $\beta$, as we will see in the next section. It is also interesting to note that in the presence of extra dimensions the imaginary part decreases and hence the perturbations of braneworld BHs are longer lived compared to their four dimensional counterpart. This is consistent with earlier findings, see e.g. \cite{Chakraborty:2017qve,Seahra:2004fg,Andriot:2017oaz}.

\section{Bound on the tidal charge from GW150914}\label{sec:bounds}
In the previous section, we (numerically) solved the perturbed gravitational field equations using the continued fraction method and determined the QNM frequencies for a fixed value of the mass and spin of a rotating braneworld BH. LIGO-Virgo GW parameter inference, on the other hand, is usually performed within a Bayesian framework. Hence, we end up with a posterior probability distribution on the mass and spin of the remnant BH. In this section, we use these publicly available LVK measurements of the mass and spin of the remnant object of GW150914~\cite{public_data_release_GWTC2_TGR} to provide a preliminary bound on the tidal charge parameter, $\beta$.

The LIGO-Virgo collaborations outlined two complementary Bayesian techniques to measure the remnant BH properties in \cite{Abbott:2020jks}. The first approach, called \pyring\ (see \cite{Carullo:2019flw,Isi:2019aib} and Section VII A.1 in~\cite{Abbott:2020jks}) infers the remnant properties by fitting a numerical relativity (NR)-inspired or a theory-agnostic damped-sinusoid ringdown template to just the post-merger signal. The outcome is a measurement of final mass and spin (or the the complex frequencies) along with additional phenomenological degrees of freedom to capture deviations from GR predictions. The final mass and spin measurements are then converted to the QNM frequencies using appropriate fitting formalae~\cite{London:2014cma,London:2018gaq}. Specifically, we use the measurements from the Kerr$_{\text{220}}$ model in~\cite{Abbott:2020jks}. The second approach, called the \pseob\ analysis (see \cite{Ghosh:2021mrv} and Section VII A.2 in~\cite{Abbott:2020jks}) attempts to make full use of the GW modelling by simultaneously measuring the inspiral and ringdown properties. Instead of using NR-inspired fitting formulae to predict the ringdown frequencies, the method leaves them as free parameters in the model and estimates them directly from the data. Both these methods are null tests of GR looking for an inconsistency with the predictions of the theory, and between them, have reported the tightest constraints on the remnant properties to date~\cite{Abbott:2020jks,Carullo:2019flw,Ghosh:2021mrv,Carullo:2021dui}. There is also a third reported measurement of the final mass and spin which indeed uses NR-inspired fitting formulae to predict the final mass and spin, starting from the masses and spins of the initial binary. These `\imr' estimates use the power in the entire signal without additional free parameters built into the model, and thus yield the tightest constraints on the measurement of $\{M_f.a_f\}$. In this paper, we treat these three methods as three independent measurements of the remnant BH properties, and check for consistency between them to obtain a preliminary bound on possible values of $\beta$ for the gravitational wave signal GW150914. We also restrict ourselves to just the least damped $\ell=2=m$ mode.

\begin{figure}
\centering
\includegraphics[scale=0.5]{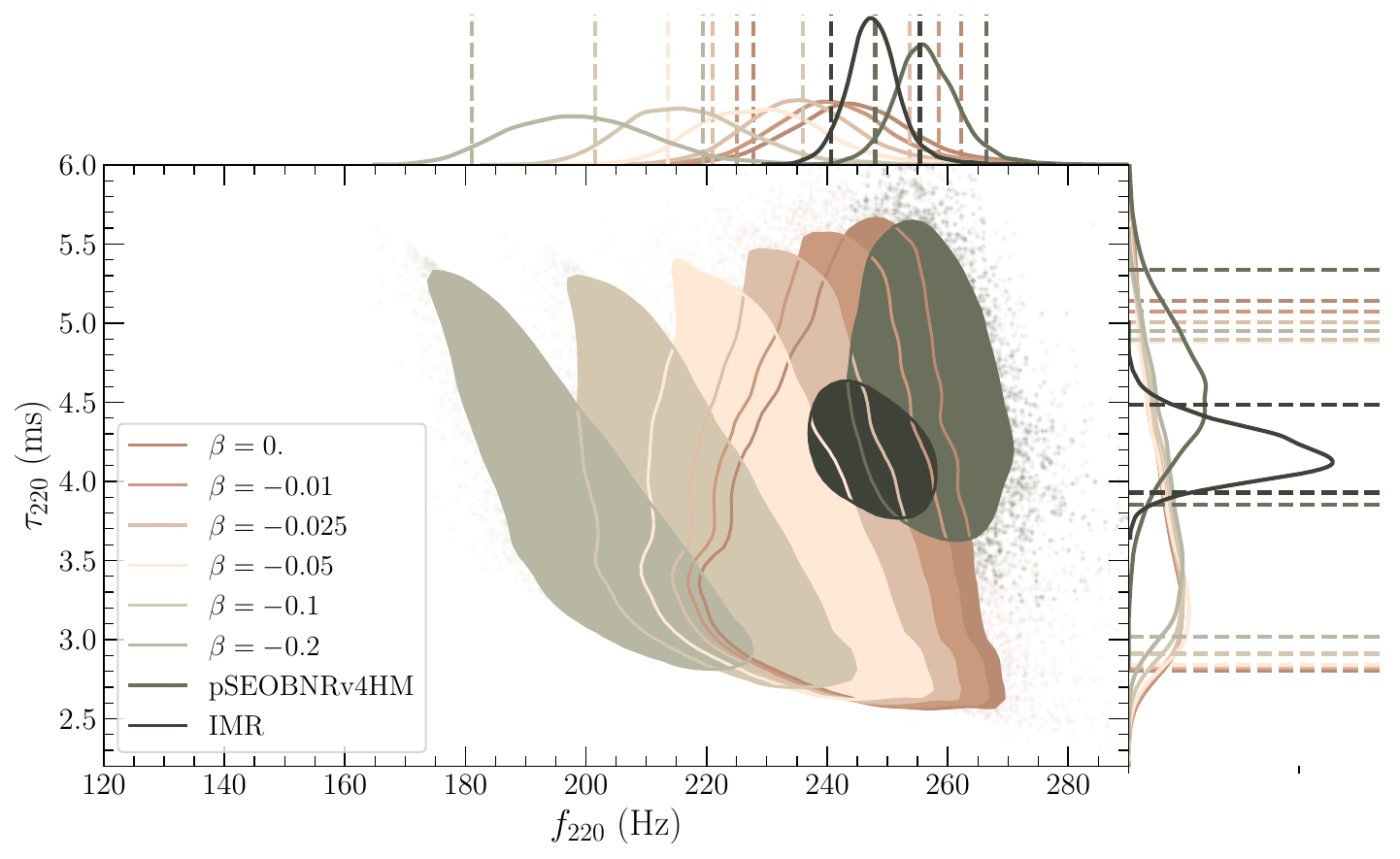}
\caption{90\% credible levels of the 2D posterior probability distributions, and the marginalised 1D posterior probability distributions (with the 90\% credible intervals) of the frequency $f_{n\ell m}$ and damping time $\tau_{n\ell m}$ of the $\ell=2=m$ mode. The \pseob\ posterior probability distribution is from~\cite{Ghosh:2021mrv} while the \imr\ posterior is from~\cite{Abbott:2020jks}. From the above posteriors, values of $\beta \lesssim -0.05$ seem to be inconsistent with the \pseob\ and \imr\ measurements.}
\label{fig_contour}
\end{figure}

Given a value for $\beta$, one can use the measured distributions of final mass and spin from the \pyring\ analysis to predict a distribution on the frequency and damping time, ($f_{220}, \tau_{220}$) (using \ref{freq_rel} and \ref{damp_rel}), and then check for their consistency with the \pseob\ and \imr\ measurements of ($f_{220},\tau_{220}$) as was reported in \cite{Ghosh:2021mrv} and \cite{Abbott:2020jks} respectively. For the case of $\beta=0$, one gets back the predictions of GR and the three distributions are consistent with each other, as shown in \ref{fig_contour} and as indeed reported in previous publications~\cite{Abbott:2020jks,Ghosh:2021mrv}. However, for non-zero values of $\beta$ one begins to find inconsistencies between the predicted and observed posteriors (\ref{fig_contour}). The inconsistencies  increase as we increase the magnitude of $\beta$. For $\beta=-0.01$ and $\beta=-0.025$, we find that the predictions of the QNM frequencies are still consistent, at the 90\% credible interval with both the \pseob\ and \imr\ measurements. But already for $\beta=-0.05$, we start seeing disagreement with the \pseob\ and \imr\ measurements. Hence, at current measurement uncertainties, values of the tidal parameter $\beta < -0.05$ appear to be unlikely.

A major caveat in the above analysis is the use of mass and spin measurements that were made assuming GR as the null hypothesis. The appropriate implementation would have been to build a complete inspiral-merger-ringdown model of GWs in the braneworld scenario, including the parameter $\beta$, and use this model to infer $(M,\chi,\beta)$ simultaneously within a Bayesian framework. Unfortunately, such a model is still some way into the future and hence, we restrict ourselves to measurements assuming GR. The uncertainties in the measurement of ($f_{220}, \tau_{220}$) in \cite{Abbott:2020jks,Carullo:2019flw,Ghosh:2021mrv} seem to also suggest that even if the GR predictions are not correct, the actual values might only vary perturbatively from them. Hence, using the GR measurements as a starting point for our analysis may be considered a safe assumption for the order-of-magnitude bounds we report on $\beta$. This also allows us to assume that $\beta$ is not correlated with the $(M,\chi)$ measurements, and hence for a given value of $\beta$, we can use the \pyring\ samples of $(M,\chi)$ to predict an ($f_{220}, \tau_{220}$) distribution. A more comprehensive study of the correlations between $(M,\chi,\beta)$ is left for future work.

\section{Discussion and Concluding Remarks}

The presence of an extra spatial dimension has distinctive signatures on the four dimensional brane, which manifest themselves in various regimes, starting from BHs to cosmology. This is because the effective gravitational field equations on the four dimensional brane gets modified by terms inherited from the higher dimensional spacetime. As a consequence, the solutions of the effective gravitational field equations on the brane differs from their GR counterparts. In the present context, for vacuum four dimensional brane spacetime, the gravitational field equations differ from the Einstein equations by an appropriate projection of the higher dimensional Weyl tensor. As a consequence, it turns out that the effective gravitational field equations resemble the Einstein-Maxwell system, with an overall negative sign for the electromagnetic stress tensor. The axisymmetric solution, arising out of this field equations looks like the Kerr-Newman spacetime, with a \emph{negative} contribution from the charge term. This drastically changes the spacetime structure, e.g., one can have the spin of such a rotating braneworld BH to be larger than unity, in striking contrast to GR. 

In this work, we have explored the implications of this charge term on the spacetime geometry through its effect on the QNM spectra of BHs. We wrote down the differential equations satisfied by the radial and angular parts of a generic spin `s' perturbation around a background rotating braneworld BH spacetime; and subsequently (numerically) solved them using the continued fraction method to obtain the (complex) QNM frequencies. These frequencies depend on the mass $M$, the dimensionless spin $\chi$ and the dimensionless tidal charge parameter $\beta$, which is negative for the braneworld scenario (while $\beta=Q^{2}/4M^{2}$, for the Kerr-Newman spacetime). The remnant object produced in the merger of two BHs is expected to \emph{ring down} into a stable final state through the emission of GWs in the form of a QNM spectra. If an extra spatial dimension is indeed present, the QNM frequencies, $(f_{n\ell m},\tau_{n\ell m})$, would be expected to depend on the tidal charge $\beta$. Hence, we try to provide a preliminary bound on possible values of $\beta$ by checking for consistency between predictions of $(f_{n\ell m},\tau_{n\ell m})$ in the braneworld scenario and publicly available measurements of the same from the LIGO-Virgo observations, for the first gravitational wave event GW150914. We find that it would be highly unlikely to have values of $\beta<-0.05$. 

The work in this paper has several possible future directions. Firstly, the constraint on the tidal charge must translate appropriately to the length of the extra dimension. This requires extending the brane solution to the bulk spacetime, which we hope to address in future work. Moreover, in this work we have used a single GW observation, namely GW150914 to impose the constraint on $\beta$, which can presumably be improved if we can combine information from multiple binary BH GW observations. Besides, modelling both the inspiral and ringdown part within the braneworld scenario would enable us to perform a full Bayesian analysis on the mass, spin and tidal charge parameter, without input from GR. We hope to address some of these questions in future work.

\section*{Acknowledgement}

The authors would like to thank everyone at the frontline of the Covid-19 pandemic. The authors are also grateful to Gregorio Carullo, Sudipta Sarkar and Soumen Roy for helpful discussions and comments. A.G. is indebted to F.Stoecker for keeping him sane during insane times. Research of AKM is supported by SERB, Government of India through the National Post Doctoral Fellowship grant (PDF/2021/003081). Research of S.C. is funded by the INSPIRE Faculty fellowship from the DST, Government of India (Reg. No. DST/INSPIRE/04/2018/000893) and by the Start-Up Research Grant from SERB, DST, Government of India (Reg. No. SRG/2020/000409). This material is based upon work supported by NSF’s LIGO Laboratory which is a major facility fully funded by the National Science Foundation.

\bibliography{Tidal_Charge_GW_v1}

\begin{thebibliography}{101}%
\makeatletter
\providecommand \@ifxundefined [1]{%
 \@ifx{#1\undefined}
}%
\providecommand \@ifnum [1]{%
 \ifnum #1\expandafter \@firstoftwo
 \else \expandafter \@secondoftwo
 \fi
}%
\providecommand \@ifx [1]{%
 \ifx #1\expandafter \@firstoftwo
 \else \expandafter \@secondoftwo
 \fi
}%
\providecommand \natexlab [1]{#1}%
\providecommand \enquote  [1]{``#1''}%
\providecommand \bibnamefont  [1]{#1}%
\providecommand \bibfnamefont [1]{#1}%
\providecommand \citenamefont [1]{#1}%
\providecommand \href@noop [0]{\@secondoftwo}%
\providecommand \href [0]{\begingroup \@sanitize@url \@href}%
\providecommand \@href[1]{\@@startlink{#1}\@@href}%
\providecommand \@@href[1]{\endgroup#1\@@endlink}%
\providecommand \@sanitize@url [0]{\catcode `\\12\catcode `\$12\catcode
  `\&12\catcode `\#12\catcode `\^12\catcode `\_12\catcode `\%12\relax}%
\providecommand \@@startlink[1]{}%
\providecommand \@@endlink[0]{}%
\providecommand \url  [0]{\begingroup\@sanitize@url \@url }%
\providecommand \@url [1]{\endgroup\@href {#1}{\urlprefix }}%
\providecommand \urlprefix  [0]{URL }%
\providecommand \Eprint [0]{\href }%
\providecommand \doibase [0]{https://doi.org/}%
\providecommand \selectlanguage [0]{\@gobble}%
\providecommand \bibinfo  [0]{\@secondoftwo}%
\providecommand \bibfield  [0]{\@secondoftwo}%
\providecommand \translation [1]{[#1]}%
\providecommand \BibitemOpen [0]{}%
\providecommand \bibitemStop [0]{}%
\providecommand \bibitemNoStop [0]{.\EOS\space}%
\providecommand \EOS [0]{\spacefactor3000\relax}%
\providecommand \BibitemShut  [1]{\csname bibitem#1\endcsname}%
\let\auto@bib@innerbib\@empty
\bibitem [{\citenamefont {Einstein}(1916)}]{Einstein:1916cc}%
  \BibitemOpen
  \bibfield  {author} {\bibinfo {author} {\bibfnamefont {A.}~\bibnamefont
  {Einstein}},\ }\bibfield  {title} {\bibinfo {title} {{Approximative
  Integration of the Field Equations of Gravitation}},\ }\href@noop {}
  {\bibfield  {journal} {\bibinfo  {journal} {Sitzungsber. Preuss. Akad. Wiss.
  Berlin (Math. Phys. )}\ }\textbf {\bibinfo {volume} {1916}},\ \bibinfo
  {pages} {688} (\bibinfo {year} {1916})}\BibitemShut {NoStop}%
\bibitem [{\citenamefont {Einstein}(1918)}]{Einstein:1918btx}%
  \BibitemOpen
  \bibfield  {author} {\bibinfo {author} {\bibfnamefont {A.}~\bibnamefont
  {Einstein}},\ }\bibfield  {title} {\bibinfo {title} {{\"Uber
  Gravitationswellen}},\ }\href@noop {} {\bibfield  {journal} {\bibinfo
  {journal} {Sitzungsber. Preuss. Akad. Wiss. Berlin (Math. Phys. )}\ }\textbf
  {\bibinfo {volume} {1918}},\ \bibinfo {pages} {154} (\bibinfo {year}
  {1918})}\BibitemShut {NoStop}%
\bibitem [{\citenamefont {Hawking}\ and\ \citenamefont
  {Ellis}(2011)}]{Hawking:1973uf}%
  \BibitemOpen
  \bibfield  {author} {\bibinfo {author} {\bibfnamefont {S.~W.}\ \bibnamefont
  {Hawking}}\ and\ \bibinfo {author} {\bibfnamefont {G.~F.~R.}\ \bibnamefont
  {Ellis}},\ }\href {https://doi.org/10.1017/CBO9780511524646} {\emph {\bibinfo
  {title} {{The Large Scale Structure of Space-Time}}}},\ Cambridge Monographs
  on Mathematical Physics\ (\bibinfo  {publisher} {Cambridge University
  Press},\ \bibinfo {year} {2011})\BibitemShut {NoStop}%
\bibitem [{\citenamefont {Wald}(1984)}]{Wald:1984rg}%
  \BibitemOpen
  \bibfield  {author} {\bibinfo {author} {\bibfnamefont {R.~M.}\ \bibnamefont
  {Wald}},\ }\href {https://doi.org/10.7208/chicago/9780226870373.001.0001}
  {\emph {\bibinfo {title} {{General Relativity}}}}\ (\bibinfo  {publisher}
  {Chicago Univ. Pr.},\ \bibinfo {address} {Chicago, USA},\ \bibinfo {year}
  {1984})\BibitemShut {NoStop}%
\bibitem [{\citenamefont {Will}(2006)}]{Will:2005va}%
  \BibitemOpen
  \bibfield  {author} {\bibinfo {author} {\bibfnamefont {C.~M.}\ \bibnamefont
  {Will}},\ }\bibfield  {title} {\bibinfo {title} {{The Confrontation between
  general relativity and experiment}},\ }\href
  {https://doi.org/10.12942/lrr-2006-3} {\bibfield  {journal} {\bibinfo
  {journal} {Living Rev. Rel.}\ }\textbf {\bibinfo {volume} {9}},\ \bibinfo
  {pages} {3} (\bibinfo {year} {2006})},\ \Eprint
  {https://arxiv.org/abs/gr-qc/0510072} {arXiv:gr-qc/0510072} \BibitemShut
  {NoStop}%
\bibitem [{\citenamefont {Berti}\ \emph {et~al.}(2015)\citenamefont {Berti}
  \emph {et~al.}}]{Berti:2015itd}%
  \BibitemOpen
  \bibfield  {author} {\bibinfo {author} {\bibfnamefont {E.}~\bibnamefont
  {Berti}} \emph {et~al.},\ }\bibfield  {title} {\bibinfo {title} {{Testing
  General Relativity with Present and Future Astrophysical Observations}},\
  }\href {https://doi.org/10.1088/0264-9381/32/24/243001} {\bibfield  {journal}
  {\bibinfo  {journal} {Class. Quant. Grav.}\ }\textbf {\bibinfo {volume}
  {32}},\ \bibinfo {pages} {243001} (\bibinfo {year} {2015})},\ \Eprint
  {https://arxiv.org/abs/1501.07274} {arXiv:1501.07274 [gr-qc]} \BibitemShut
  {NoStop}%
\bibitem [{\citenamefont {Berti}\ \emph
  {et~al.}(2018{\natexlab{a}})\citenamefont {Berti}, \citenamefont {Yagi},\
  and\ \citenamefont {Yunes}}]{Berti:2018cxi}%
  \BibitemOpen
  \bibfield  {author} {\bibinfo {author} {\bibfnamefont {E.}~\bibnamefont
  {Berti}}, \bibinfo {author} {\bibfnamefont {K.}~\bibnamefont {Yagi}},\ and\
  \bibinfo {author} {\bibfnamefont {N.}~\bibnamefont {Yunes}},\ }\bibfield
  {title} {\bibinfo {title} {{Extreme Gravity Tests with Gravitational Waves
  from Compact Binary Coalescences: (I) Inspiral-Merger}},\ }\href
  {https://doi.org/10.1007/s10714-018-2362-8} {\bibfield  {journal} {\bibinfo
  {journal} {Gen. Rel. Grav.}\ }\textbf {\bibinfo {volume} {50}},\ \bibinfo
  {pages} {46} (\bibinfo {year} {2018}{\natexlab{a}})},\ \Eprint
  {https://arxiv.org/abs/1801.03208} {arXiv:1801.03208 [gr-qc]} \BibitemShut
  {NoStop}%
\bibitem [{\citenamefont {Berti}\ \emph
  {et~al.}(2018{\natexlab{b}})\citenamefont {Berti}, \citenamefont {Yagi},
  \citenamefont {Yang},\ and\ \citenamefont {Yunes}}]{Berti:2018vdi}%
  \BibitemOpen
  \bibfield  {author} {\bibinfo {author} {\bibfnamefont {E.}~\bibnamefont
  {Berti}}, \bibinfo {author} {\bibfnamefont {K.}~\bibnamefont {Yagi}},
  \bibinfo {author} {\bibfnamefont {H.}~\bibnamefont {Yang}},\ and\ \bibinfo
  {author} {\bibfnamefont {N.}~\bibnamefont {Yunes}},\ }\bibfield  {title}
  {\bibinfo {title} {{Extreme Gravity Tests with Gravitational Waves from
  Compact Binary Coalescences: (II) Ringdown}},\ }\href
  {https://doi.org/10.1007/s10714-018-2372-6} {\bibfield  {journal} {\bibinfo
  {journal} {Gen. Rel. Grav.}\ }\textbf {\bibinfo {volume} {50}},\ \bibinfo
  {pages} {49} (\bibinfo {year} {2018}{\natexlab{b}})},\ \Eprint
  {https://arxiv.org/abs/1801.03587} {arXiv:1801.03587 [gr-qc]} \BibitemShut
  {NoStop}%
\bibitem [{\citenamefont {Abbott}\ \emph
  {et~al.}(2019{\natexlab{a}})\citenamefont {Abbott} \emph
  {et~al.}}]{LIGOScientific:2019fpa}%
  \BibitemOpen
  \bibfield  {author} {\bibinfo {author} {\bibfnamefont {B.~P.}\ \bibnamefont
  {Abbott}} \emph {et~al.} (\bibinfo {collaboration} {LIGO Scientific,
  Virgo}),\ }\bibfield  {title} {\bibinfo {title} {{Tests of General Relativity
  with the Binary Black Hole Signals from the LIGO-Virgo Catalog GWTC-1}},\
  }\href {https://doi.org/10.1103/PhysRevD.100.104036} {\bibfield  {journal}
  {\bibinfo  {journal} {Phys. Rev. D}\ }\textbf {\bibinfo {volume} {100}},\
  \bibinfo {pages} {104036} (\bibinfo {year} {2019}{\natexlab{a}})},\ \Eprint
  {https://arxiv.org/abs/1903.04467} {arXiv:1903.04467 [gr-qc]} \BibitemShut
  {NoStop}%
\bibitem [{\citenamefont {Abbott}\ \emph
  {et~al.}(2020{\natexlab{a}})\citenamefont {Abbott} \emph
  {et~al.}}]{Abbott:2020jks}%
  \BibitemOpen
  \bibfield  {author} {\bibinfo {author} {\bibfnamefont {R.}~\bibnamefont
  {Abbott}} \emph {et~al.} (\bibinfo {collaboration} {LIGO Scientific,
  Virgo}),\ }\bibfield  {title} {\bibinfo {title} {{Tests of General Relativity
  with Binary Black Holes from the second LIGO-Virgo Gravitational-Wave
  Transient Catalog}},\ }\href@noop {} {\  (\bibinfo {year}
  {2020}{\natexlab{a}})},\ \Eprint {https://arxiv.org/abs/2010.14529}
  {arXiv:2010.14529 [gr-qc]} \BibitemShut {NoStop}%
\bibitem [{\citenamefont {Abbott}\ \emph {et~al.}(2017)\citenamefont {Abbott}
  \emph {et~al.}}]{TheLIGOScientific:2017qsa}%
  \BibitemOpen
  \bibfield  {author} {\bibinfo {author} {\bibfnamefont {B.~P.}\ \bibnamefont
  {Abbott}} \emph {et~al.} (\bibinfo {collaboration} {LIGO Scientific,
  Virgo}),\ }\bibfield  {title} {\bibinfo {title} {{GW170817: Observation of
  Gravitational Waves from a Binary Neutron Star Inspiral}},\ }\href
  {https://doi.org/10.1103/PhysRevLett.119.161101} {\bibfield  {journal}
  {\bibinfo  {journal} {Phys. Rev. Lett.}\ }\textbf {\bibinfo {volume} {119}},\
  \bibinfo {pages} {161101} (\bibinfo {year} {2017})},\ \Eprint
  {https://arxiv.org/abs/1710.05832} {arXiv:1710.05832 [gr-qc]} \BibitemShut
  {NoStop}%
\bibitem [{\citenamefont {Akiyama}\ \emph {et~al.}(2019)\citenamefont {Akiyama}
  \emph {et~al.}}]{Akiyama:2019cqa}%
  \BibitemOpen
  \bibfield  {author} {\bibinfo {author} {\bibfnamefont {K.}~\bibnamefont
  {Akiyama}} \emph {et~al.} (\bibinfo {collaboration} {Event Horizon
  Telescope}),\ }\bibfield  {title} {\bibinfo {title} {{First M87 Event Horizon
  Telescope Results. I. The Shadow of the Supermassive Black Hole}},\ }\href
  {https://doi.org/10.3847/2041-8213/ab0ec7} {\bibfield  {journal} {\bibinfo
  {journal} {Astrophys. J. Lett.}\ }\textbf {\bibinfo {volume} {875}},\
  \bibinfo {pages} {L1} (\bibinfo {year} {2019})},\ \Eprint
  {https://arxiv.org/abs/1906.11238} {arXiv:1906.11238 [astro-ph.GA]}
  \BibitemShut {NoStop}%
\bibitem [{\citenamefont {Birrell}\ and\ \citenamefont
  {Davies}(1984)}]{Birrell:1982ix}%
  \BibitemOpen
  \bibfield  {author} {\bibinfo {author} {\bibfnamefont {N.}~\bibnamefont
  {Birrell}}\ and\ \bibinfo {author} {\bibfnamefont {P.}~\bibnamefont
  {Davies}},\ }\href {https://doi.org/10.1017/CBO9780511622632} {\emph
  {\bibinfo {title} {{Quantum Fields in Curved Space}}}},\ Cambridge Monographs
  on Mathematical Physics\ (\bibinfo  {publisher} {Cambridge Univ. Press},\
  \bibinfo {address} {Cambridge, UK},\ \bibinfo {year} {1984})\BibitemShut
  {NoStop}%
\bibitem [{\citenamefont {Penrose}(1965)}]{PhysRevLett.14.57}%
  \BibitemOpen
  \bibfield  {author} {\bibinfo {author} {\bibfnamefont {R.}~\bibnamefont
  {Penrose}},\ }\bibfield  {title} {\bibinfo {title} {Gravitational collapse
  and space-time singularities},\ }\href
  {https://doi.org/10.1103/PhysRevLett.14.57} {\bibfield  {journal} {\bibinfo
  {journal} {Phys. Rev. Lett.}\ }\textbf {\bibinfo {volume} {14}},\ \bibinfo
  {pages} {57} (\bibinfo {year} {1965})}\BibitemShut {NoStop}%
\bibitem [{\citenamefont {Hawking}(1976)}]{PhysRevD.14.2460}%
  \BibitemOpen
  \bibfield  {author} {\bibinfo {author} {\bibfnamefont {S.~W.}\ \bibnamefont
  {Hawking}},\ }\bibfield  {title} {\bibinfo {title} {Breakdown of
  predictability in gravitational collapse},\ }\href
  {https://doi.org/10.1103/PhysRevD.14.2460} {\bibfield  {journal} {\bibinfo
  {journal} {Phys. Rev. D}\ }\textbf {\bibinfo {volume} {14}},\ \bibinfo
  {pages} {2460} (\bibinfo {year} {1976})}\BibitemShut {NoStop}%
\bibitem [{\citenamefont {Cardoso}\ \emph {et~al.}(2018)\citenamefont
  {Cardoso}, \citenamefont {Costa}, \citenamefont {Destounis}, \citenamefont
  {Hintz},\ and\ \citenamefont {Jansen}}]{Cardoso:2017soq}%
  \BibitemOpen
  \bibfield  {author} {\bibinfo {author} {\bibfnamefont {V.}~\bibnamefont
  {Cardoso}}, \bibinfo {author} {\bibfnamefont {J.~a.~L.}\ \bibnamefont
  {Costa}}, \bibinfo {author} {\bibfnamefont {K.}~\bibnamefont {Destounis}},
  \bibinfo {author} {\bibfnamefont {P.}~\bibnamefont {Hintz}},\ and\ \bibinfo
  {author} {\bibfnamefont {A.}~\bibnamefont {Jansen}},\ }\bibfield  {title}
  {\bibinfo {title} {{Quasinormal modes and Strong Cosmic Censorship}},\ }\href
  {https://doi.org/10.1103/PhysRevLett.120.031103} {\bibfield  {journal}
  {\bibinfo  {journal} {Phys. Rev. Lett.}\ }\textbf {\bibinfo {volume} {120}},\
  \bibinfo {pages} {031103} (\bibinfo {year} {2018})},\ \Eprint
  {https://arxiv.org/abs/1711.10502} {arXiv:1711.10502 [gr-qc]} \BibitemShut
  {NoStop}%
\bibitem [{\citenamefont {Rahman}\ \emph {et~al.}(2019)\citenamefont {Rahman},
  \citenamefont {Chakraborty}, \citenamefont {SenGupta},\ and\ \citenamefont
  {Sen}}]{Rahman:2018oso}%
  \BibitemOpen
  \bibfield  {author} {\bibinfo {author} {\bibfnamefont {M.}~\bibnamefont
  {Rahman}}, \bibinfo {author} {\bibfnamefont {S.}~\bibnamefont {Chakraborty}},
  \bibinfo {author} {\bibfnamefont {S.}~\bibnamefont {SenGupta}},\ and\
  \bibinfo {author} {\bibfnamefont {A.~A.}\ \bibnamefont {Sen}},\ }\bibfield
  {title} {\bibinfo {title} {{Fate of Strong Cosmic Censorship Conjecture in
  Presence of Higher Spacetime Dimensions}},\ }\href
  {https://doi.org/10.1007/JHEP03(2019)178} {\bibfield  {journal} {\bibinfo
  {journal} {JHEP}\ }\textbf {\bibinfo {volume} {03}},\ \bibinfo {pages}
  {178}},\ \Eprint {https://arxiv.org/abs/1811.08538} {arXiv:1811.08538
  [gr-qc]} \BibitemShut {NoStop}%
\bibitem [{\citenamefont {Weinberg}(1989)}]{RevModPhys.61.1}%
  \BibitemOpen
  \bibfield  {author} {\bibinfo {author} {\bibfnamefont {S.}~\bibnamefont
  {Weinberg}},\ }\bibfield  {title} {\bibinfo {title} {The cosmological
  constant problem},\ }\href {https://doi.org/10.1103/RevModPhys.61.1}
  {\bibfield  {journal} {\bibinfo  {journal} {Rev. Mod. Phys.}\ }\textbf
  {\bibinfo {volume} {61}},\ \bibinfo {pages} {1} (\bibinfo {year}
  {1989})}\BibitemShut {NoStop}%
\bibitem [{\citenamefont {Padmanabhan}(2003)}]{Padmanabhan:2002ji}%
  \BibitemOpen
  \bibfield  {author} {\bibinfo {author} {\bibfnamefont {T.}~\bibnamefont
  {Padmanabhan}},\ }\bibfield  {title} {\bibinfo {title} {{Cosmological
  constant: The Weight of the vacuum}},\ }\href
  {https://doi.org/10.1016/S0370-1573(03)00120-0} {\bibfield  {journal}
  {\bibinfo  {journal} {Phys. Rept.}\ }\textbf {\bibinfo {volume} {380}},\
  \bibinfo {pages} {235} (\bibinfo {year} {2003})},\ \Eprint
  {https://arxiv.org/abs/hep-th/0212290} {arXiv:hep-th/0212290} \BibitemShut
  {NoStop}%
\bibitem [{\citenamefont {Carroll}(2001)}]{Carroll:2000fy}%
  \BibitemOpen
  \bibfield  {author} {\bibinfo {author} {\bibfnamefont {S.~M.}\ \bibnamefont
  {Carroll}},\ }\bibfield  {title} {\bibinfo {title} {{The Cosmological
  constant}},\ }\href {https://doi.org/10.12942/lrr-2001-1} {\bibfield
  {journal} {\bibinfo  {journal} {Living Rev. Rel.}\ }\textbf {\bibinfo
  {volume} {4}},\ \bibinfo {pages} {1} (\bibinfo {year} {2001})},\ \Eprint
  {https://arxiv.org/abs/astro-ph/0004075} {arXiv:astro-ph/0004075}
  \BibitemShut {NoStop}%
\bibitem [{\citenamefont {Clifton}\ \emph {et~al.}(2012)\citenamefont
  {Clifton}, \citenamefont {Ferreira}, \citenamefont {Padilla},\ and\
  \citenamefont {Skordis}}]{Clifton:2011jh}%
  \BibitemOpen
  \bibfield  {author} {\bibinfo {author} {\bibfnamefont {T.}~\bibnamefont
  {Clifton}}, \bibinfo {author} {\bibfnamefont {P.~G.}\ \bibnamefont
  {Ferreira}}, \bibinfo {author} {\bibfnamefont {A.}~\bibnamefont {Padilla}},\
  and\ \bibinfo {author} {\bibfnamefont {C.}~\bibnamefont {Skordis}},\
  }\bibfield  {title} {\bibinfo {title} {{Modified Gravity and Cosmology}},\
  }\href {https://doi.org/10.1016/j.physrep.2012.01.001} {\bibfield  {journal}
  {\bibinfo  {journal} {Phys. Rept.}\ }\textbf {\bibinfo {volume} {513}},\
  \bibinfo {pages} {1} (\bibinfo {year} {2012})},\ \Eprint
  {https://arxiv.org/abs/1106.2476} {arXiv:1106.2476 [astro-ph.CO]}
  \BibitemShut {NoStop}%
\bibitem [{\citenamefont {Capozziello}\ and\ \citenamefont
  {De~Laurentis}(2011)}]{Capozziello:2011et}%
  \BibitemOpen
  \bibfield  {author} {\bibinfo {author} {\bibfnamefont {S.}~\bibnamefont
  {Capozziello}}\ and\ \bibinfo {author} {\bibfnamefont {M.}~\bibnamefont
  {De~Laurentis}},\ }\bibfield  {title} {\bibinfo {title} {{Extended Theories
  of Gravity}},\ }\href {https://doi.org/10.1016/j.physrep.2011.09.003}
  {\bibfield  {journal} {\bibinfo  {journal} {Phys. Rept.}\ }\textbf {\bibinfo
  {volume} {509}},\ \bibinfo {pages} {167} (\bibinfo {year} {2011})},\ \Eprint
  {https://arxiv.org/abs/1108.6266} {arXiv:1108.6266 [gr-qc]} \BibitemShut
  {NoStop}%
\bibitem [{\citenamefont {Nojiri}\ and\ \citenamefont
  {Odintsov}(2011)}]{Nojiri:2010wj}%
  \BibitemOpen
  \bibfield  {author} {\bibinfo {author} {\bibfnamefont {S.}~\bibnamefont
  {Nojiri}}\ and\ \bibinfo {author} {\bibfnamefont {S.~D.}\ \bibnamefont
  {Odintsov}},\ }\bibfield  {title} {\bibinfo {title} {{Unified cosmic history
  in modified gravity: from F(R) theory to Lorentz non-invariant models}},\
  }\href {https://doi.org/10.1016/j.physrep.2011.04.001} {\bibfield  {journal}
  {\bibinfo  {journal} {Phys. Rept.}\ }\textbf {\bibinfo {volume} {505}},\
  \bibinfo {pages} {59} (\bibinfo {year} {2011})},\ \Eprint
  {https://arxiv.org/abs/1011.0544} {arXiv:1011.0544 [gr-qc]} \BibitemShut
  {NoStop}%
\bibitem [{\citenamefont {Gleyzes}\ \emph {et~al.}(2015)\citenamefont
  {Gleyzes}, \citenamefont {Langlois}, \citenamefont {Piazza},\ and\
  \citenamefont {Vernizzi}}]{Gleyzes:2014dya}%
  \BibitemOpen
  \bibfield  {author} {\bibinfo {author} {\bibfnamefont {J.}~\bibnamefont
  {Gleyzes}}, \bibinfo {author} {\bibfnamefont {D.}~\bibnamefont {Langlois}},
  \bibinfo {author} {\bibfnamefont {F.}~\bibnamefont {Piazza}},\ and\ \bibinfo
  {author} {\bibfnamefont {F.}~\bibnamefont {Vernizzi}},\ }\bibfield  {title}
  {\bibinfo {title} {{Healthy theories beyond Horndeski}},\ }\href
  {https://doi.org/10.1103/PhysRevLett.114.211101} {\bibfield  {journal}
  {\bibinfo  {journal} {Phys. Rev. Lett.}\ }\textbf {\bibinfo {volume} {114}},\
  \bibinfo {pages} {211101} (\bibinfo {year} {2015})},\ \Eprint
  {https://arxiv.org/abs/1404.6495} {arXiv:1404.6495 [hep-th]} \BibitemShut
  {NoStop}%
\bibitem [{\citenamefont {Woodard}(2015)}]{Woodard:2015zca}%
  \BibitemOpen
  \bibfield  {author} {\bibinfo {author} {\bibfnamefont {R.~P.}\ \bibnamefont
  {Woodard}},\ }\bibfield  {title} {\bibinfo {title} {{Ostrogradsky's theorem
  on Hamiltonian instability}},\ }\href
  {https://doi.org/10.4249/scholarpedia.32243} {\bibfield  {journal} {\bibinfo
  {journal} {Scholarpedia}\ }\textbf {\bibinfo {volume} {10}},\ \bibinfo
  {pages} {32243} (\bibinfo {year} {2015})},\ \Eprint
  {https://arxiv.org/abs/1506.02210} {arXiv:1506.02210 [hep-th]} \BibitemShut
  {NoStop}%
\bibitem [{\citenamefont {Abbott}\ \emph
  {et~al.}(2019{\natexlab{b}})\citenamefont {Abbott} \emph
  {et~al.}}]{LIGOScientific:2018mvr}%
  \BibitemOpen
  \bibfield  {author} {\bibinfo {author} {\bibfnamefont {B.~P.}\ \bibnamefont
  {Abbott}} \emph {et~al.} (\bibinfo {collaboration} {LIGO Scientific,
  Virgo}),\ }\bibfield  {title} {\bibinfo {title} {{GWTC-1: A
  Gravitational-Wave Transient Catalog of Compact Binary Mergers Observed by
  LIGO and Virgo during the First and Second Observing Runs}},\ }\href
  {https://doi.org/10.1103/PhysRevX.9.031040} {\bibfield  {journal} {\bibinfo
  {journal} {Phys. Rev. X}\ }\textbf {\bibinfo {volume} {9}},\ \bibinfo {pages}
  {031040} (\bibinfo {year} {2019}{\natexlab{b}})},\ \Eprint
  {https://arxiv.org/abs/1811.12907} {arXiv:1811.12907 [astro-ph.HE]}
  \BibitemShut {NoStop}%
\bibitem [{\citenamefont {Abbott}\ \emph
  {et~al.}(2020{\natexlab{b}})\citenamefont {Abbott} \emph
  {et~al.}}]{Abbott:2020niy}%
  \BibitemOpen
  \bibfield  {author} {\bibinfo {author} {\bibfnamefont {R.}~\bibnamefont
  {Abbott}} \emph {et~al.} (\bibinfo {collaboration} {LIGO Scientific,
  Virgo}),\ }\bibfield  {title} {\bibinfo {title} {{GWTC-2: Compact Binary
  Coalescences Observed by LIGO and Virgo During the First Half of the Third
  Observing Run}},\ }\href@noop {} {\  (\bibinfo {year}
  {2020}{\natexlab{b}})},\ \Eprint {https://arxiv.org/abs/2010.14527}
  {arXiv:2010.14527 [gr-qc]} \BibitemShut {NoStop}%
\bibitem [{\citenamefont {Aasi}\ \emph {et~al.}(2015)\citenamefont {Aasi} \emph
  {et~al.}}]{TheLIGOScientific:2014jea}%
  \BibitemOpen
  \bibfield  {author} {\bibinfo {author} {\bibfnamefont {J.}~\bibnamefont
  {Aasi}} \emph {et~al.} (\bibinfo {collaboration} {LIGO Scientific}),\
  }\bibfield  {title} {\bibinfo {title} {{Advanced LIGO}},\ }\href
  {https://doi.org/10.1088/0264-9381/32/7/074001} {\bibfield  {journal}
  {\bibinfo  {journal} {Class. Quant. Grav.}\ }\textbf {\bibinfo {volume}
  {32}},\ \bibinfo {pages} {074001} (\bibinfo {year} {2015})},\ \Eprint
  {https://arxiv.org/abs/1411.4547} {arXiv:1411.4547 [gr-qc]} \BibitemShut
  {NoStop}%
\bibitem [{\citenamefont {Acernese}\ \emph {et~al.}(2015)\citenamefont
  {Acernese} \emph {et~al.}}]{TheVirgo:2014hva}%
  \BibitemOpen
  \bibfield  {author} {\bibinfo {author} {\bibfnamefont {F.}~\bibnamefont
  {Acernese}} \emph {et~al.} (\bibinfo {collaboration} {VIRGO}),\ }\bibfield
  {title} {\bibinfo {title} {{Advanced Virgo: a second-generation
  interferometric gravitational wave detector}},\ }\href
  {https://doi.org/10.1088/0264-9381/32/2/024001} {\bibfield  {journal}
  {\bibinfo  {journal} {Class. Quant. Grav.}\ }\textbf {\bibinfo {volume}
  {32}},\ \bibinfo {pages} {024001} (\bibinfo {year} {2015})},\ \Eprint
  {https://arxiv.org/abs/1408.3978} {arXiv:1408.3978 [gr-qc]} \BibitemShut
  {NoStop}%
\bibitem [{\citenamefont {Baker}\ \emph {et~al.}(2017)\citenamefont {Baker},
  \citenamefont {Bellini}, \citenamefont {Ferreira}, \citenamefont {Lagos},
  \citenamefont {Noller},\ and\ \citenamefont {Sawicki}}]{Baker:2017hug}%
  \BibitemOpen
  \bibfield  {author} {\bibinfo {author} {\bibfnamefont {T.}~\bibnamefont
  {Baker}}, \bibinfo {author} {\bibfnamefont {E.}~\bibnamefont {Bellini}},
  \bibinfo {author} {\bibfnamefont {P.~G.}\ \bibnamefont {Ferreira}}, \bibinfo
  {author} {\bibfnamefont {M.}~\bibnamefont {Lagos}}, \bibinfo {author}
  {\bibfnamefont {J.}~\bibnamefont {Noller}},\ and\ \bibinfo {author}
  {\bibfnamefont {I.}~\bibnamefont {Sawicki}},\ }\bibfield  {title} {\bibinfo
  {title} {{Strong constraints on cosmological gravity from GW170817 and GRB
  170817A}},\ }\href {https://doi.org/10.1103/PhysRevLett.119.251301}
  {\bibfield  {journal} {\bibinfo  {journal} {Phys. Rev. Lett.}\ }\textbf
  {\bibinfo {volume} {119}},\ \bibinfo {pages} {251301} (\bibinfo {year}
  {2017})},\ \Eprint {https://arxiv.org/abs/1710.06394} {arXiv:1710.06394
  [astro-ph.CO]} \BibitemShut {NoStop}%
\bibitem [{\citenamefont {Jana}\ \emph {et~al.}(2018)\citenamefont {Jana},
  \citenamefont {Chakravarty},\ and\ \citenamefont {Mohanty}}]{Jana:2017ost}%
  \BibitemOpen
  \bibfield  {author} {\bibinfo {author} {\bibfnamefont {S.}~\bibnamefont
  {Jana}}, \bibinfo {author} {\bibfnamefont {G.~K.}\ \bibnamefont
  {Chakravarty}},\ and\ \bibinfo {author} {\bibfnamefont {S.}~\bibnamefont
  {Mohanty}},\ }\bibfield  {title} {\bibinfo {title} {{Constraints on
  Born-Infeld gravity from the speed of gravitational waves after GW170817 and
  GRB 170817A}},\ }\href {https://doi.org/10.1103/PhysRevD.97.084011}
  {\bibfield  {journal} {\bibinfo  {journal} {Phys. Rev. D}\ }\textbf {\bibinfo
  {volume} {97}},\ \bibinfo {pages} {084011} (\bibinfo {year} {2018})},\
  \Eprint {https://arxiv.org/abs/1711.04137} {arXiv:1711.04137 [gr-qc]}
  \BibitemShut {NoStop}%
\bibitem [{\citenamefont {Creminelli}\ and\ \citenamefont
  {Vernizzi}(2017)}]{Creminelli:2017sry}%
  \BibitemOpen
  \bibfield  {author} {\bibinfo {author} {\bibfnamefont {P.}~\bibnamefont
  {Creminelli}}\ and\ \bibinfo {author} {\bibfnamefont {F.}~\bibnamefont
  {Vernizzi}},\ }\bibfield  {title} {\bibinfo {title} {{Dark Energy after
  GW170817 and GRB170817A}},\ }\href
  {https://doi.org/10.1103/PhysRevLett.119.251302} {\bibfield  {journal}
  {\bibinfo  {journal} {Phys. Rev. Lett.}\ }\textbf {\bibinfo {volume} {119}},\
  \bibinfo {pages} {251302} (\bibinfo {year} {2017})},\ \Eprint
  {https://arxiv.org/abs/1710.05877} {arXiv:1710.05877 [astro-ph.CO]}
  \BibitemShut {NoStop}%
\bibitem [{\citenamefont {Nojiri}\ and\ \citenamefont
  {Odintsov}(2018)}]{Nojiri:2017hai}%
  \BibitemOpen
  \bibfield  {author} {\bibinfo {author} {\bibfnamefont {S.}~\bibnamefont
  {Nojiri}}\ and\ \bibinfo {author} {\bibfnamefont {S.~D.}\ \bibnamefont
  {Odintsov}},\ }\bibfield  {title} {\bibinfo {title} {{Cosmological Bound from
  the Neutron Star Merger GW170817 in scalar-tensor and $F(R)$ gravity
  theories}},\ }\href {https://doi.org/10.1016/j.physletb.2018.01.078}
  {\bibfield  {journal} {\bibinfo  {journal} {Phys. Lett. B}\ }\textbf
  {\bibinfo {volume} {779}},\ \bibinfo {pages} {425} (\bibinfo {year}
  {2018})},\ \Eprint {https://arxiv.org/abs/1711.00492} {arXiv:1711.00492
  [astro-ph.CO]} \BibitemShut {NoStop}%
\bibitem [{\citenamefont {Akrami}\ \emph {et~al.}(2018)\citenamefont {Akrami},
  \citenamefont {Brax}, \citenamefont {Davis},\ and\ \citenamefont
  {Vardanyan}}]{Akrami:2018yjz}%
  \BibitemOpen
  \bibfield  {author} {\bibinfo {author} {\bibfnamefont {Y.}~\bibnamefont
  {Akrami}}, \bibinfo {author} {\bibfnamefont {P.}~\bibnamefont {Brax}},
  \bibinfo {author} {\bibfnamefont {A.-C.}\ \bibnamefont {Davis}},\ and\
  \bibinfo {author} {\bibfnamefont {V.}~\bibnamefont {Vardanyan}},\ }\bibfield
  {title} {\bibinfo {title} {{Neutron star merger GW170817 strongly constrains
  doubly coupled bigravity}},\ }\href
  {https://doi.org/10.1103/PhysRevD.97.124010} {\bibfield  {journal} {\bibinfo
  {journal} {Phys. Rev. D}\ }\textbf {\bibinfo {volume} {97}},\ \bibinfo
  {pages} {124010} (\bibinfo {year} {2018})},\ \Eprint
  {https://arxiv.org/abs/1803.09726} {arXiv:1803.09726 [astro-ph.CO]}
  \BibitemShut {NoStop}%
\bibitem [{\citenamefont {Chakravarti}\ \emph {et~al.}(2020)\citenamefont
  {Chakravarti}, \citenamefont {Chakraborty}, \citenamefont {Phukon},
  \citenamefont {Bose},\ and\ \citenamefont {SenGupta}}]{Chakravarti:2019aup}%
  \BibitemOpen
  \bibfield  {author} {\bibinfo {author} {\bibfnamefont {K.}~\bibnamefont
  {Chakravarti}}, \bibinfo {author} {\bibfnamefont {S.}~\bibnamefont
  {Chakraborty}}, \bibinfo {author} {\bibfnamefont {K.~S.}\ \bibnamefont
  {Phukon}}, \bibinfo {author} {\bibfnamefont {S.}~\bibnamefont {Bose}},\ and\
  \bibinfo {author} {\bibfnamefont {S.}~\bibnamefont {SenGupta}},\ }\bibfield
  {title} {\bibinfo {title} {{Constraining extra-spatial dimensions with
  observations of GW170817}},\ }\href
  {https://doi.org/10.1088/1361-6382/ab8355} {\bibfield  {journal} {\bibinfo
  {journal} {Class. Quant. Grav.}\ }\textbf {\bibinfo {volume} {37}},\ \bibinfo
  {pages} {105004} (\bibinfo {year} {2020})},\ \Eprint
  {https://arxiv.org/abs/1903.10159} {arXiv:1903.10159 [gr-qc]} \BibitemShut
  {NoStop}%
\bibitem [{\citenamefont {Ghosh}\ \emph {et~al.}(2019)\citenamefont {Ghosh},
  \citenamefont {Jana}, \citenamefont {Mishra},\ and\ \citenamefont
  {Sarkar}}]{Ghosh:2019twk}%
  \BibitemOpen
  \bibfield  {author} {\bibinfo {author} {\bibfnamefont {A.}~\bibnamefont
  {Ghosh}}, \bibinfo {author} {\bibfnamefont {S.}~\bibnamefont {Jana}},
  \bibinfo {author} {\bibfnamefont {A.~K.}\ \bibnamefont {Mishra}},\ and\
  \bibinfo {author} {\bibfnamefont {S.}~\bibnamefont {Sarkar}},\ }\bibfield
  {title} {\bibinfo {title} {{Constraints on higher curvature gravity from time
  delay between GW170817 and GRB 170817A}},\ }\href
  {https://doi.org/10.1103/PhysRevD.100.084054} {\bibfield  {journal} {\bibinfo
   {journal} {Phys. Rev. D}\ }\textbf {\bibinfo {volume} {100}},\ \bibinfo
  {pages} {084054} (\bibinfo {year} {2019})},\ \Eprint
  {https://arxiv.org/abs/1906.08014} {arXiv:1906.08014 [gr-qc]} \BibitemShut
  {NoStop}%
\bibitem [{\citenamefont {Yu}\ \emph {et~al.}(2017)\citenamefont {Yu},
  \citenamefont {Gu}, \citenamefont {Huang}, \citenamefont {Wang},
  \citenamefont {Meng},\ and\ \citenamefont {Liu}}]{Yu:2016tar}%
  \BibitemOpen
  \bibfield  {author} {\bibinfo {author} {\bibfnamefont {H.}~\bibnamefont
  {Yu}}, \bibinfo {author} {\bibfnamefont {B.-M.}\ \bibnamefont {Gu}}, \bibinfo
  {author} {\bibfnamefont {F.~P.}\ \bibnamefont {Huang}}, \bibinfo {author}
  {\bibfnamefont {Y.-Q.}\ \bibnamefont {Wang}}, \bibinfo {author}
  {\bibfnamefont {X.-H.}\ \bibnamefont {Meng}},\ and\ \bibinfo {author}
  {\bibfnamefont {Y.-X.}\ \bibnamefont {Liu}},\ }\bibfield  {title} {\bibinfo
  {title} {{Probing extra dimension through gravitational wave observations of
  compact binaries and their electromagnetic counterparts}},\ }\href
  {https://doi.org/10.1088/1475-7516/2017/02/039} {\bibfield  {journal}
  {\bibinfo  {journal} {JCAP}\ }\textbf {\bibinfo {volume} {02}},\ \bibinfo
  {pages} {039}},\ \Eprint {https://arxiv.org/abs/1607.03388} {arXiv:1607.03388
  [gr-qc]} \BibitemShut {NoStop}%
\bibitem [{\citenamefont {Lin}\ \emph {et~al.}(2020)\citenamefont {Lin},
  \citenamefont {Yu},\ and\ \citenamefont {Liu}}]{Lin:2020wnp}%
  \BibitemOpen
  \bibfield  {author} {\bibinfo {author} {\bibfnamefont {Z.-C.}\ \bibnamefont
  {Lin}}, \bibinfo {author} {\bibfnamefont {H.}~\bibnamefont {Yu}},\ and\
  \bibinfo {author} {\bibfnamefont {Y.-X.}\ \bibnamefont {Liu}},\ }\bibfield
  {title} {\bibinfo {title} {{Constraint on the radius of five-dimensional dS
  spacetime with GW170817 and GRB 170817A}},\ }\href
  {https://doi.org/10.1103/PhysRevD.101.104058} {\bibfield  {journal} {\bibinfo
   {journal} {Phys. Rev. D}\ }\textbf {\bibinfo {volume} {101}},\ \bibinfo
  {pages} {104058} (\bibinfo {year} {2020})},\ \Eprint
  {https://arxiv.org/abs/2001.06581} {arXiv:2001.06581 [gr-qc]} \BibitemShut
  {NoStop}%
\bibitem [{\citenamefont {Wang}\ \emph {et~al.}(2021)\citenamefont {Wang},
  \citenamefont {Tang}, \citenamefont {Li},\ and\ \citenamefont
  {Fan}}]{Wang:2021uuh}%
  \BibitemOpen
  \bibfield  {author} {\bibinfo {author} {\bibfnamefont {H.-T.}\ \bibnamefont
  {Wang}}, \bibinfo {author} {\bibfnamefont {S.-P.}\ \bibnamefont {Tang}},
  \bibinfo {author} {\bibfnamefont {P.-C.}\ \bibnamefont {Li}},\ and\ \bibinfo
  {author} {\bibfnamefont {Y.-Z.}\ \bibnamefont {Fan}},\ }\bibfield  {title}
  {\bibinfo {title} {{Quasinormal-modes of the Kerr-Newman black hole: GW150914
  and fundamental physics implications}},\ }\href@noop {} {\  (\bibinfo {year}
  {2021})},\ \Eprint {https://arxiv.org/abs/2104.07594} {arXiv:2104.07594
  [gr-qc]} \BibitemShut {NoStop}%
\bibitem [{\citenamefont {Psaltis}\ \emph {et~al.}(2020)\citenamefont {Psaltis}
  \emph {et~al.}}]{Psaltis:2020lvx}%
  \BibitemOpen
  \bibfield  {author} {\bibinfo {author} {\bibfnamefont {D.}~\bibnamefont
  {Psaltis}} \emph {et~al.} (\bibinfo {collaboration} {Event Horizon
  Telescope}),\ }\bibfield  {title} {\bibinfo {title} {{Gravitational Test
  Beyond the First Post-Newtonian Order with the Shadow of the M87 Black
  Hole}},\ }\href {https://doi.org/10.1103/PhysRevLett.125.141104} {\bibfield
  {journal} {\bibinfo  {journal} {Phys. Rev. Lett.}\ }\textbf {\bibinfo
  {volume} {125}},\ \bibinfo {pages} {141104} (\bibinfo {year} {2020})},\
  \Eprint {https://arxiv.org/abs/2010.01055} {arXiv:2010.01055 [gr-qc]}
  \BibitemShut {NoStop}%
\bibitem [{\citenamefont {Zhu}\ \emph {et~al.}(2018)\citenamefont {Zhu},
  \citenamefont {Johnson},\ and\ \citenamefont {Narayan}}]{Zhu_2018}%
  \BibitemOpen
  \bibfield  {author} {\bibinfo {author} {\bibfnamefont {Z.}~\bibnamefont
  {Zhu}}, \bibinfo {author} {\bibfnamefont {M.~D.}\ \bibnamefont {Johnson}},\
  and\ \bibinfo {author} {\bibfnamefont {R.}~\bibnamefont {Narayan}},\
  }\bibfield  {title} {\bibinfo {title} {Testing general relativity with the
  black hole shadow size and asymmetry of sagittarius a*: Limitations from
  interstellar scattering},\ }\href {https://doi.org/10.3847/1538-4357/aaef3d}
  {\bibfield  {journal} {\bibinfo  {journal} {The Astrophysical Journal}\
  }\textbf {\bibinfo {volume} {870}},\ \bibinfo {pages} {6} (\bibinfo {year}
  {2018})}\BibitemShut {NoStop}%
\bibitem [{\citenamefont {Banerjee}\ \emph {et~al.}(2020)\citenamefont
  {Banerjee}, \citenamefont {Chakraborty},\ and\ \citenamefont
  {SenGupta}}]{Banerjee:2019nnj}%
  \BibitemOpen
  \bibfield  {author} {\bibinfo {author} {\bibfnamefont {I.}~\bibnamefont
  {Banerjee}}, \bibinfo {author} {\bibfnamefont {S.}~\bibnamefont
  {Chakraborty}},\ and\ \bibinfo {author} {\bibfnamefont {S.}~\bibnamefont
  {SenGupta}},\ }\bibfield  {title} {\bibinfo {title} {{Silhouette of M87*: A
  New Window to Peek into the World of Hidden Dimensions}},\ }\href
  {https://doi.org/10.1103/PhysRevD.101.041301} {\bibfield  {journal} {\bibinfo
   {journal} {Phys. Rev. D}\ }\textbf {\bibinfo {volume} {101}},\ \bibinfo
  {pages} {041301} (\bibinfo {year} {2020})},\ \Eprint
  {https://arxiv.org/abs/1909.09385} {arXiv:1909.09385 [gr-qc]} \BibitemShut
  {NoStop}%
\bibitem [{\citenamefont {Maartens}(2004)}]{Maartens:2003tw}%
  \BibitemOpen
  \bibfield  {author} {\bibinfo {author} {\bibfnamefont {R.}~\bibnamefont
  {Maartens}},\ }\bibfield  {title} {\bibinfo {title} {{Brane world gravity}},\
  }\href {https://doi.org/10.12942/lrr-2004-7} {\bibfield  {journal} {\bibinfo
  {journal} {Living Rev. Rel.}\ }\textbf {\bibinfo {volume} {7}},\ \bibinfo
  {pages} {7} (\bibinfo {year} {2004})},\ \Eprint
  {https://arxiv.org/abs/gr-qc/0312059} {arXiv:gr-qc/0312059} \BibitemShut
  {NoStop}%
\bibitem [{\citenamefont {Csaki}(2004)}]{Csaki:2004ay}%
  \BibitemOpen
  \bibfield  {author} {\bibinfo {author} {\bibfnamefont {C.}~\bibnamefont
  {Csaki}},\ }\bibfield  {title} {\bibinfo {title} {{TASI lectures on extra
  dimensions and branes}},\ }in\ \href@noop {} {\emph {\bibinfo {booktitle}
  {{Theoretical Advanced Study Institute in Elementary Particle Physics (TASI
  2002): Particle Physics and Cosmology: The Quest for Physics Beyond the
  Standard Model(s)}}}}\ (\bibinfo {year} {2004})\ \Eprint
  {https://arxiv.org/abs/hep-ph/0404096} {arXiv:hep-ph/0404096} \BibitemShut
  {NoStop}%
\bibitem [{\citenamefont {Perez-Lorenzana}(2005)}]{PerezLorenzana:2005iv}%
  \BibitemOpen
  \bibfield  {author} {\bibinfo {author} {\bibfnamefont {A.}~\bibnamefont
  {Perez-Lorenzana}},\ }\bibfield  {title} {\bibinfo {title} {{An Introduction
  to extra dimensions}},\ }\href {https://doi.org/10.1088/1742-6596/18/1/006}
  {\bibfield  {journal} {\bibinfo  {journal} {J. Phys. Conf. Ser.}\ }\textbf
  {\bibinfo {volume} {18}},\ \bibinfo {pages} {224} (\bibinfo {year} {2005})},\
  \Eprint {https://arxiv.org/abs/hep-ph/0503177} {arXiv:hep-ph/0503177}
  \BibitemShut {NoStop}%
\bibitem [{\citenamefont {Kanti}(2004)}]{Kanti:2004nr}%
  \BibitemOpen
  \bibfield  {author} {\bibinfo {author} {\bibfnamefont {P.}~\bibnamefont
  {Kanti}},\ }\bibfield  {title} {\bibinfo {title} {{Black holes in theories
  with large extra dimensions: A Review}},\ }\href
  {https://doi.org/10.1142/S0217751X04018324} {\bibfield  {journal} {\bibinfo
  {journal} {Int. J. Mod. Phys. A}\ }\textbf {\bibinfo {volume} {19}},\
  \bibinfo {pages} {4899} (\bibinfo {year} {2004})},\ \Eprint
  {https://arxiv.org/abs/hep-ph/0402168} {arXiv:hep-ph/0402168} \BibitemShut
  {NoStop}%
\bibitem [{\citenamefont {Overduin}\ and\ \citenamefont
  {Wesson}(1997)}]{Overduin:1998pn}%
  \BibitemOpen
  \bibfield  {author} {\bibinfo {author} {\bibfnamefont {J.~M.}\ \bibnamefont
  {Overduin}}\ and\ \bibinfo {author} {\bibfnamefont {P.~S.}\ \bibnamefont
  {Wesson}},\ }\bibfield  {title} {\bibinfo {title} {{Kaluza-Klein gravity}},\
  }\href {https://doi.org/10.1016/S0370-1573(96)00046-4} {\bibfield  {journal}
  {\bibinfo  {journal} {Phys. Rept.}\ }\textbf {\bibinfo {volume} {283}},\
  \bibinfo {pages} {303} (\bibinfo {year} {1997})},\ \Eprint
  {https://arxiv.org/abs/gr-qc/9805018} {arXiv:gr-qc/9805018} \BibitemShut
  {NoStop}%
\bibitem [{\citenamefont {Antoniadis}\ \emph {et~al.}(1998)\citenamefont
  {Antoniadis}, \citenamefont {Arkani-Hamed}, \citenamefont {Dimopoulos},\ and\
  \citenamefont {Dvali}}]{Antoniadis:1998ig}%
  \BibitemOpen
  \bibfield  {author} {\bibinfo {author} {\bibfnamefont {I.}~\bibnamefont
  {Antoniadis}}, \bibinfo {author} {\bibfnamefont {N.}~\bibnamefont
  {Arkani-Hamed}}, \bibinfo {author} {\bibfnamefont {S.}~\bibnamefont
  {Dimopoulos}},\ and\ \bibinfo {author} {\bibfnamefont {G.~R.}\ \bibnamefont
  {Dvali}},\ }\bibfield  {title} {\bibinfo {title} {{New dimensions at a
  millimeter to a Fermi and superstrings at a TeV}},\ }\href
  {https://doi.org/10.1016/S0370-2693(98)00860-0} {\bibfield  {journal}
  {\bibinfo  {journal} {Phys. Lett. B}\ }\textbf {\bibinfo {volume} {436}},\
  \bibinfo {pages} {257} (\bibinfo {year} {1998})},\ \Eprint
  {https://arxiv.org/abs/hep-ph/9804398} {arXiv:hep-ph/9804398} \BibitemShut
  {NoStop}%
\bibitem [{\citenamefont {Randall}\ and\ \citenamefont
  {Sundrum}(1999{\natexlab{a}})}]{Randall:1999ee}%
  \BibitemOpen
  \bibfield  {author} {\bibinfo {author} {\bibfnamefont {L.}~\bibnamefont
  {Randall}}\ and\ \bibinfo {author} {\bibfnamefont {R.}~\bibnamefont
  {Sundrum}},\ }\bibfield  {title} {\bibinfo {title} {{A Large mass hierarchy
  from a small extra dimension}},\ }\href
  {https://doi.org/10.1103/PhysRevLett.83.3370} {\bibfield  {journal} {\bibinfo
   {journal} {Phys. Rev. Lett.}\ }\textbf {\bibinfo {volume} {83}},\ \bibinfo
  {pages} {3370} (\bibinfo {year} {1999}{\natexlab{a}})},\ \Eprint
  {https://arxiv.org/abs/hep-ph/9905221} {arXiv:hep-ph/9905221} \BibitemShut
  {NoStop}%
\bibitem [{\citenamefont {Aad}\ \emph {et~al.}(2012)\citenamefont {Aad} \emph
  {et~al.}}]{Aad:2012tfa}%
  \BibitemOpen
  \bibfield  {author} {\bibinfo {author} {\bibfnamefont {G.}~\bibnamefont
  {Aad}} \emph {et~al.} (\bibinfo {collaboration} {ATLAS}),\ }\bibfield
  {title} {\bibinfo {title} {{Observation of a new particle in the search for
  the Standard Model Higgs boson with the ATLAS detector at the LHC}},\ }\href
  {https://doi.org/10.1016/j.physletb.2012.08.020} {\bibfield  {journal}
  {\bibinfo  {journal} {Phys. Lett. B}\ }\textbf {\bibinfo {volume} {716}},\
  \bibinfo {pages} {1} (\bibinfo {year} {2012})},\ \Eprint
  {https://arxiv.org/abs/1207.7214} {arXiv:1207.7214 [hep-ex]} \BibitemShut
  {NoStop}%
\bibitem [{\citenamefont {Chatrchyan}\ \emph {et~al.}(2012)\citenamefont
  {Chatrchyan} \emph {et~al.}}]{Chatrchyan:2012ufa}%
  \BibitemOpen
  \bibfield  {author} {\bibinfo {author} {\bibfnamefont {S.}~\bibnamefont
  {Chatrchyan}} \emph {et~al.} (\bibinfo {collaboration} {CMS}),\ }\bibfield
  {title} {\bibinfo {title} {{Observation of a New Boson at a Mass of 125 GeV
  with the CMS Experiment at the LHC}},\ }\href
  {https://doi.org/10.1016/j.physletb.2012.08.021} {\bibfield  {journal}
  {\bibinfo  {journal} {Phys. Lett. B}\ }\textbf {\bibinfo {volume} {716}},\
  \bibinfo {pages} {30} (\bibinfo {year} {2012})},\ \Eprint
  {https://arxiv.org/abs/1207.7235} {arXiv:1207.7235 [hep-ex]} \BibitemShut
  {NoStop}%
\bibitem [{\citenamefont {Arkani-Hamed}\ \emph {et~al.}(1998)\citenamefont
  {Arkani-Hamed}, \citenamefont {Dimopoulos},\ and\ \citenamefont
  {Dvali}}]{ArkaniHamed:1998rs}%
  \BibitemOpen
  \bibfield  {author} {\bibinfo {author} {\bibfnamefont {N.}~\bibnamefont
  {Arkani-Hamed}}, \bibinfo {author} {\bibfnamefont {S.}~\bibnamefont
  {Dimopoulos}},\ and\ \bibinfo {author} {\bibfnamefont {G.~R.}\ \bibnamefont
  {Dvali}},\ }\bibfield  {title} {\bibinfo {title} {{The Hierarchy problem and
  new dimensions at a millimeter}},\ }\href
  {https://doi.org/10.1016/S0370-2693(98)00466-3} {\bibfield  {journal}
  {\bibinfo  {journal} {Phys. Lett. B}\ }\textbf {\bibinfo {volume} {429}},\
  \bibinfo {pages} {263} (\bibinfo {year} {1998})},\ \Eprint
  {https://arxiv.org/abs/hep-ph/9803315} {arXiv:hep-ph/9803315} \BibitemShut
  {NoStop}%
\bibitem [{\citenamefont {Randall}\ and\ \citenamefont
  {Sundrum}(1999{\natexlab{b}})}]{Randall:1999vf}%
  \BibitemOpen
  \bibfield  {author} {\bibinfo {author} {\bibfnamefont {L.}~\bibnamefont
  {Randall}}\ and\ \bibinfo {author} {\bibfnamefont {R.}~\bibnamefont
  {Sundrum}},\ }\bibfield  {title} {\bibinfo {title} {{An Alternative to
  compactification}},\ }\href {https://doi.org/10.1103/PhysRevLett.83.4690}
  {\bibfield  {journal} {\bibinfo  {journal} {Phys. Rev. Lett.}\ }\textbf
  {\bibinfo {volume} {83}},\ \bibinfo {pages} {4690} (\bibinfo {year}
  {1999}{\natexlab{b}})},\ \Eprint {https://arxiv.org/abs/hep-th/9906064}
  {arXiv:hep-th/9906064} \BibitemShut {NoStop}%
\bibitem [{\citenamefont {Harko}\ and\ \citenamefont
  {Mak}(2004)}]{Harko:2004ui}%
  \BibitemOpen
  \bibfield  {author} {\bibinfo {author} {\bibfnamefont {T.}~\bibnamefont
  {Harko}}\ and\ \bibinfo {author} {\bibfnamefont {M.~K.}\ \bibnamefont
  {Mak}},\ }\bibfield  {title} {\bibinfo {title} {{Vacuum solutions of the
  gravitational field equations in the brane world model}},\ }\href
  {https://doi.org/10.1103/PhysRevD.69.064020} {\bibfield  {journal} {\bibinfo
  {journal} {Phys. Rev. D}\ }\textbf {\bibinfo {volume} {69}},\ \bibinfo
  {pages} {064020} (\bibinfo {year} {2004})},\ \Eprint
  {https://arxiv.org/abs/gr-qc/0401049} {arXiv:gr-qc/0401049} \BibitemShut
  {NoStop}%
\bibitem [{\citenamefont {Aliev}\ and\ \citenamefont
  {Gumrukcuoglu}(2005)}]{Aliev:2005bi}%
  \BibitemOpen
  \bibfield  {author} {\bibinfo {author} {\bibfnamefont {A.~N.}\ \bibnamefont
  {Aliev}}\ and\ \bibinfo {author} {\bibfnamefont {A.~E.}\ \bibnamefont
  {Gumrukcuoglu}},\ }\bibfield  {title} {\bibinfo {title} {{Charged rotating
  black holes on a 3-brane}},\ }\href
  {https://doi.org/10.1103/PhysRevD.71.104027} {\bibfield  {journal} {\bibinfo
  {journal} {Phys. Rev. D}\ }\textbf {\bibinfo {volume} {71}},\ \bibinfo
  {pages} {104027} (\bibinfo {year} {2005})},\ \Eprint
  {https://arxiv.org/abs/hep-th/0502223} {arXiv:hep-th/0502223} \BibitemShut
  {NoStop}%
\bibitem [{\citenamefont {Chakraborty}\ and\ \citenamefont
  {SenGupta}(2016)}]{Chakraborty:2015taq}%
  \BibitemOpen
  \bibfield  {author} {\bibinfo {author} {\bibfnamefont {S.}~\bibnamefont
  {Chakraborty}}\ and\ \bibinfo {author} {\bibfnamefont {S.}~\bibnamefont
  {SenGupta}},\ }\bibfield  {title} {\bibinfo {title} {{Spherically symmetric
  brane in a bulk of $f(R)$ and Gauss\textendash{}Bonnet gravity}},\ }\href
  {https://doi.org/10.1088/0264-9381/33/22/225001} {\bibfield  {journal}
  {\bibinfo  {journal} {Class. Quant. Grav.}\ }\textbf {\bibinfo {volume}
  {33}},\ \bibinfo {pages} {225001} (\bibinfo {year} {2016})},\ \Eprint
  {https://arxiv.org/abs/1510.01953} {arXiv:1510.01953 [gr-qc]} \BibitemShut
  {NoStop}%
\bibitem [{\citenamefont {Chakraborty}\ and\ \citenamefont
  {SenGupta}(2015)}]{Chakraborty:2014xla}%
  \BibitemOpen
  \bibfield  {author} {\bibinfo {author} {\bibfnamefont {S.}~\bibnamefont
  {Chakraborty}}\ and\ \bibinfo {author} {\bibfnamefont {S.}~\bibnamefont
  {SenGupta}},\ }\bibfield  {title} {\bibinfo {title} {{Spherically symmetric
  brane spacetime with bulk $f(\mathcal {R})$ gravity}},\ }\href
  {https://doi.org/10.1140/epjc/s10052-014-3234-3} {\bibfield  {journal}
  {\bibinfo  {journal} {Eur. Phys. J. C}\ }\textbf {\bibinfo {volume} {75}},\
  \bibinfo {pages} {11} (\bibinfo {year} {2015})},\ \Eprint
  {https://arxiv.org/abs/1409.4115} {arXiv:1409.4115 [gr-qc]} \BibitemShut
  {NoStop}%
\bibitem [{\citenamefont {Dadhich}\ \emph {et~al.}(2000)\citenamefont
  {Dadhich}, \citenamefont {Maartens}, \citenamefont {Papadopoulos},\ and\
  \citenamefont {Rezania}}]{Dadhich:2000am}%
  \BibitemOpen
  \bibfield  {author} {\bibinfo {author} {\bibfnamefont {N.}~\bibnamefont
  {Dadhich}}, \bibinfo {author} {\bibfnamefont {R.}~\bibnamefont {Maartens}},
  \bibinfo {author} {\bibfnamefont {P.}~\bibnamefont {Papadopoulos}},\ and\
  \bibinfo {author} {\bibfnamefont {V.}~\bibnamefont {Rezania}},\ }\bibfield
  {title} {\bibinfo {title} {{Black holes on the brane}},\ }\href
  {https://doi.org/10.1016/S0370-2693(00)00798-X} {\bibfield  {journal}
  {\bibinfo  {journal} {Phys. Lett. B}\ }\textbf {\bibinfo {volume} {487}},\
  \bibinfo {pages} {1} (\bibinfo {year} {2000})},\ \Eprint
  {https://arxiv.org/abs/hep-th/0003061} {arXiv:hep-th/0003061} \BibitemShut
  {NoStop}%
\bibitem [{\citenamefont {Chamblin}\ \emph {et~al.}(2000)\citenamefont
  {Chamblin}, \citenamefont {Hawking},\ and\ \citenamefont
  {Reall}}]{Chamblin:1999by}%
  \BibitemOpen
  \bibfield  {author} {\bibinfo {author} {\bibfnamefont {A.}~\bibnamefont
  {Chamblin}}, \bibinfo {author} {\bibfnamefont {S.~W.}\ \bibnamefont
  {Hawking}},\ and\ \bibinfo {author} {\bibfnamefont {H.~S.}\ \bibnamefont
  {Reall}},\ }\bibfield  {title} {\bibinfo {title} {{Brane world black
  holes}},\ }\href {https://doi.org/10.1103/PhysRevD.61.065007} {\bibfield
  {journal} {\bibinfo  {journal} {Phys. Rev. D}\ }\textbf {\bibinfo {volume}
  {61}},\ \bibinfo {pages} {065007} (\bibinfo {year} {2000})},\ \Eprint
  {https://arxiv.org/abs/hep-th/9909205} {arXiv:hep-th/9909205} \BibitemShut
  {NoStop}%
\bibitem [{\citenamefont {Chamblin}\ \emph {et~al.}(2001)\citenamefont
  {Chamblin}, \citenamefont {Reall}, \citenamefont {Shinkai},\ and\
  \citenamefont {Shiromizu}}]{Chamblin:2000ra}%
  \BibitemOpen
  \bibfield  {author} {\bibinfo {author} {\bibfnamefont {A.}~\bibnamefont
  {Chamblin}}, \bibinfo {author} {\bibfnamefont {H.~S.}\ \bibnamefont {Reall}},
  \bibinfo {author} {\bibfnamefont {H.-a.}\ \bibnamefont {Shinkai}},\ and\
  \bibinfo {author} {\bibfnamefont {T.}~\bibnamefont {Shiromizu}},\ }\bibfield
  {title} {\bibinfo {title} {{Charged brane world black holes}},\ }\href
  {https://doi.org/10.1103/PhysRevD.63.064015} {\bibfield  {journal} {\bibinfo
  {journal} {Phys. Rev. D}\ }\textbf {\bibinfo {volume} {63}},\ \bibinfo
  {pages} {064015} (\bibinfo {year} {2001})},\ \Eprint
  {https://arxiv.org/abs/hep-th/0008177} {arXiv:hep-th/0008177} \BibitemShut
  {NoStop}%
\bibitem [{\citenamefont {Emparan}\ \emph {et~al.}(2000)\citenamefont
  {Emparan}, \citenamefont {Horowitz},\ and\ \citenamefont
  {Myers}}]{Emparan:1999wa}%
  \BibitemOpen
  \bibfield  {author} {\bibinfo {author} {\bibfnamefont {R.}~\bibnamefont
  {Emparan}}, \bibinfo {author} {\bibfnamefont {G.~T.}\ \bibnamefont
  {Horowitz}},\ and\ \bibinfo {author} {\bibfnamefont {R.~C.}\ \bibnamefont
  {Myers}},\ }\bibfield  {title} {\bibinfo {title} {{Exact description of black
  holes on branes}},\ }\href {https://doi.org/10.1088/1126-6708/2000/01/007}
  {\bibfield  {journal} {\bibinfo  {journal} {JHEP}\ }\textbf {\bibinfo
  {volume} {01}},\ \bibinfo {pages} {007}},\ \Eprint
  {https://arxiv.org/abs/hep-th/9911043} {arXiv:hep-th/9911043} \BibitemShut
  {NoStop}%
\bibitem [{\citenamefont {Nakas}\ and\ \citenamefont
  {Kanti}(2021{\natexlab{a}})}]{Nakas:2020sey}%
  \BibitemOpen
  \bibfield  {author} {\bibinfo {author} {\bibfnamefont {T.}~\bibnamefont
  {Nakas}}\ and\ \bibinfo {author} {\bibfnamefont {P.}~\bibnamefont {Kanti}},\
  }\bibfield  {title} {\bibinfo {title} {{Localized brane-world black hole
  analytically connected to an AdS$_5$ boundary}},\ }\href
  {https://doi.org/10.1016/j.physletb.2021.136278} {\bibfield  {journal}
  {\bibinfo  {journal} {Phys. Lett. B}\ }\textbf {\bibinfo {volume} {816}},\
  \bibinfo {pages} {136278} (\bibinfo {year} {2021}{\natexlab{a}})},\ \Eprint
  {https://arxiv.org/abs/2012.09199} {arXiv:2012.09199 [hep-th]} \BibitemShut
  {NoStop}%
\bibitem [{\citenamefont {Nakas}\ and\ \citenamefont
  {Kanti}(2021{\natexlab{b}})}]{Nakas:2021srr}%
  \BibitemOpen
  \bibfield  {author} {\bibinfo {author} {\bibfnamefont {T.}~\bibnamefont
  {Nakas}}\ and\ \bibinfo {author} {\bibfnamefont {P.}~\bibnamefont {Kanti}},\
  }\bibfield  {title} {\bibinfo {title} {{Analytic and exponentially localized
  brane-world Reissner-Nordstr\"om-AdS solution: a top-down approach}},\
  }\href@noop {} {\  (\bibinfo {year} {2021}{\natexlab{b}})},\ \Eprint
  {https://arxiv.org/abs/2105.06915} {arXiv:2105.06915 [hep-th]} \BibitemShut
  {NoStop}%
\bibitem [{\citenamefont {Csaki}\ \emph {et~al.}(2000)\citenamefont {Csaki},
  \citenamefont {Graesser}, \citenamefont {Randall},\ and\ \citenamefont
  {Terning}}]{Csaki:1999mp}%
  \BibitemOpen
  \bibfield  {author} {\bibinfo {author} {\bibfnamefont {C.}~\bibnamefont
  {Csaki}}, \bibinfo {author} {\bibfnamefont {M.}~\bibnamefont {Graesser}},
  \bibinfo {author} {\bibfnamefont {L.}~\bibnamefont {Randall}},\ and\ \bibinfo
  {author} {\bibfnamefont {J.}~\bibnamefont {Terning}},\ }\bibfield  {title}
  {\bibinfo {title} {{Cosmology of brane models with radion stabilization}},\
  }\href {https://doi.org/10.1103/PhysRevD.62.045015} {\bibfield  {journal}
  {\bibinfo  {journal} {Phys. Rev. D}\ }\textbf {\bibinfo {volume} {62}},\
  \bibinfo {pages} {045015} (\bibinfo {year} {2000})},\ \Eprint
  {https://arxiv.org/abs/hep-ph/9911406} {arXiv:hep-ph/9911406} \BibitemShut
  {NoStop}%
\bibitem [{\citenamefont {Csaki}\ \emph {et~al.}(1999)\citenamefont {Csaki},
  \citenamefont {Graesser}, \citenamefont {Kolda},\ and\ \citenamefont
  {Terning}}]{Csaki:1999jh}%
  \BibitemOpen
  \bibfield  {author} {\bibinfo {author} {\bibfnamefont {C.}~\bibnamefont
  {Csaki}}, \bibinfo {author} {\bibfnamefont {M.}~\bibnamefont {Graesser}},
  \bibinfo {author} {\bibfnamefont {C.~F.}\ \bibnamefont {Kolda}},\ and\
  \bibinfo {author} {\bibfnamefont {J.}~\bibnamefont {Terning}},\ }\bibfield
  {title} {\bibinfo {title} {{Cosmology of one extra dimension with localized
  gravity}},\ }\href {https://doi.org/10.1016/S0370-2693(99)00896-5} {\bibfield
   {journal} {\bibinfo  {journal} {Phys. Lett. B}\ }\textbf {\bibinfo {volume}
  {462}},\ \bibinfo {pages} {34} (\bibinfo {year} {1999})},\ \Eprint
  {https://arxiv.org/abs/hep-ph/9906513} {arXiv:hep-ph/9906513} \BibitemShut
  {NoStop}%
\bibitem [{\citenamefont {Abbott}\ \emph
  {et~al.}(2019{\natexlab{c}})\citenamefont {Abbott} \emph
  {et~al.}}]{Abbott:2018lct}%
  \BibitemOpen
  \bibfield  {author} {\bibinfo {author} {\bibfnamefont {B.~P.}\ \bibnamefont
  {Abbott}} \emph {et~al.} (\bibinfo {collaboration} {LIGO Scientific,
  Virgo}),\ }\bibfield  {title} {\bibinfo {title} {{Tests of General Relativity
  with GW170817}},\ }\href {https://doi.org/10.1103/PhysRevLett.123.011102}
  {\bibfield  {journal} {\bibinfo  {journal} {Phys. Rev. Lett.}\ }\textbf
  {\bibinfo {volume} {123}},\ \bibinfo {pages} {011102} (\bibinfo {year}
  {2019}{\natexlab{c}})},\ \Eprint {https://arxiv.org/abs/1811.00364}
  {arXiv:1811.00364 [gr-qc]} \BibitemShut {NoStop}%
\bibitem [{\citenamefont {Visinelli}\ \emph {et~al.}(2018)\citenamefont
  {Visinelli}, \citenamefont {Bolis},\ and\ \citenamefont
  {Vagnozzi}}]{Visinelli:2017bny}%
  \BibitemOpen
  \bibfield  {author} {\bibinfo {author} {\bibfnamefont {L.}~\bibnamefont
  {Visinelli}}, \bibinfo {author} {\bibfnamefont {N.}~\bibnamefont {Bolis}},\
  and\ \bibinfo {author} {\bibfnamefont {S.}~\bibnamefont {Vagnozzi}},\
  }\bibfield  {title} {\bibinfo {title} {{Brane-world extra dimensions in light
  of GW170817}},\ }\href {https://doi.org/10.1103/PhysRevD.97.064039}
  {\bibfield  {journal} {\bibinfo  {journal} {Phys. Rev. D}\ }\textbf {\bibinfo
  {volume} {97}},\ \bibinfo {pages} {064039} (\bibinfo {year} {2018})},\
  \Eprint {https://arxiv.org/abs/1711.06628} {arXiv:1711.06628 [gr-qc]}
  \BibitemShut {NoStop}%
\bibitem [{\citenamefont {Chakraborty}\ \emph {et~al.}(2018)\citenamefont
  {Chakraborty}, \citenamefont {Chakravarti}, \citenamefont {Bose},\ and\
  \citenamefont {SenGupta}}]{Chakraborty:2017qve}%
  \BibitemOpen
  \bibfield  {author} {\bibinfo {author} {\bibfnamefont {S.}~\bibnamefont
  {Chakraborty}}, \bibinfo {author} {\bibfnamefont {K.}~\bibnamefont
  {Chakravarti}}, \bibinfo {author} {\bibfnamefont {S.}~\bibnamefont {Bose}},\
  and\ \bibinfo {author} {\bibfnamefont {S.}~\bibnamefont {SenGupta}},\
  }\bibfield  {title} {\bibinfo {title} {{Signatures of extra dimensions in
  gravitational waves from black hole quasinormal modes}},\ }\href
  {https://doi.org/10.1103/PhysRevD.97.104053} {\bibfield  {journal} {\bibinfo
  {journal} {Phys. Rev. D}\ }\textbf {\bibinfo {volume} {97}},\ \bibinfo
  {pages} {104053} (\bibinfo {year} {2018})},\ \Eprint
  {https://arxiv.org/abs/1710.05188} {arXiv:1710.05188 [gr-qc]} \BibitemShut
  {NoStop}%
\bibitem [{\citenamefont {Chakravarti}\ \emph {et~al.}(2019)\citenamefont
  {Chakravarti}, \citenamefont {Chakraborty}, \citenamefont {Bose},\ and\
  \citenamefont {SenGupta}}]{Chakravarti:2018vlt}%
  \BibitemOpen
  \bibfield  {author} {\bibinfo {author} {\bibfnamefont {K.}~\bibnamefont
  {Chakravarti}}, \bibinfo {author} {\bibfnamefont {S.}~\bibnamefont
  {Chakraborty}}, \bibinfo {author} {\bibfnamefont {S.}~\bibnamefont {Bose}},\
  and\ \bibinfo {author} {\bibfnamefont {S.}~\bibnamefont {SenGupta}},\
  }\bibfield  {title} {\bibinfo {title} {{Tidal Love numbers of black holes and
  neutron stars in the presence of higher dimensions: Implications of
  GW170817}},\ }\href {https://doi.org/10.1103/PhysRevD.99.024036} {\bibfield
  {journal} {\bibinfo  {journal} {Phys. Rev. D}\ }\textbf {\bibinfo {volume}
  {99}},\ \bibinfo {pages} {024036} (\bibinfo {year} {2019})},\ \Eprint
  {https://arxiv.org/abs/1811.11364} {arXiv:1811.11364 [gr-qc]} \BibitemShut
  {NoStop}%
\bibitem [{\citenamefont {Toshmatov}\ \emph {et~al.}(2016)\citenamefont
  {Toshmatov}, \citenamefont {Stuchl\'\i{}k}, \citenamefont {Schee},\ and\
  \citenamefont {Ahmedov}}]{Toshmatov:2016bsb}%
  \BibitemOpen
  \bibfield  {author} {\bibinfo {author} {\bibfnamefont {B.}~\bibnamefont
  {Toshmatov}}, \bibinfo {author} {\bibfnamefont {Z.}~\bibnamefont
  {Stuchl\'\i{}k}}, \bibinfo {author} {\bibfnamefont {J.}~\bibnamefont
  {Schee}},\ and\ \bibinfo {author} {\bibfnamefont {B.}~\bibnamefont
  {Ahmedov}},\ }\bibfield  {title} {\bibinfo {title} {{Quasinormal frequencies
  of black hole in the braneworld}},\ }\href
  {https://doi.org/10.1103/PhysRevD.93.124017} {\bibfield  {journal} {\bibinfo
  {journal} {Phys. Rev. D}\ }\textbf {\bibinfo {volume} {93}},\ \bibinfo
  {pages} {124017} (\bibinfo {year} {2016})},\ \Eprint
  {https://arxiv.org/abs/1605.02058} {arXiv:1605.02058 [gr-qc]} \BibitemShut
  {NoStop}%
\bibitem [{\citenamefont {Dey}\ \emph {et~al.}(2020{\natexlab{a}})\citenamefont
  {Dey}, \citenamefont {Biswas},\ and\ \citenamefont
  {Chakraborty}}]{Dey:2020pth}%
  \BibitemOpen
  \bibfield  {author} {\bibinfo {author} {\bibfnamefont {R.}~\bibnamefont
  {Dey}}, \bibinfo {author} {\bibfnamefont {S.}~\bibnamefont {Biswas}},\ and\
  \bibinfo {author} {\bibfnamefont {S.}~\bibnamefont {Chakraborty}},\
  }\bibfield  {title} {\bibinfo {title} {{Ergoregion instability and echoes for
  braneworld black holes: Scalar, electromagnetic and gravitational
  perturbations}},\ }\href@noop {} {\  (\bibinfo {year}
  {2020}{\natexlab{a}})},\ \Eprint {https://arxiv.org/abs/2010.07966}
  {arXiv:2010.07966 [gr-qc]} \BibitemShut {NoStop}%
\bibitem [{\citenamefont {de~Oliveira}(2020)}]{deOliveira:2020lzp}%
  \BibitemOpen
  \bibfield  {author} {\bibinfo {author} {\bibfnamefont {E.~S.}\ \bibnamefont
  {de~Oliveira}},\ }\bibfield  {title} {\bibinfo {title} {{Tidal-charge effects
  on the superradiance of rotating black holes}},\ }\href
  {https://doi.org/10.1140/epjc/s10052-020-08570-y} {\bibfield  {journal}
  {\bibinfo  {journal} {Eur. Phys. J. C}\ }\textbf {\bibinfo {volume} {80}},\
  \bibinfo {pages} {1048} (\bibinfo {year} {2020})},\ \Eprint
  {https://arxiv.org/abs/2004.10122} {arXiv:2004.10122 [gr-qc]} \BibitemShut
  {NoStop}%
\bibitem [{\citenamefont {Dey}\ \emph {et~al.}(2020{\natexlab{b}})\citenamefont
  {Dey}, \citenamefont {Chakraborty},\ and\ \citenamefont
  {Afshordi}}]{Dey:2020lhq}%
  \BibitemOpen
  \bibfield  {author} {\bibinfo {author} {\bibfnamefont {R.}~\bibnamefont
  {Dey}}, \bibinfo {author} {\bibfnamefont {S.}~\bibnamefont {Chakraborty}},\
  and\ \bibinfo {author} {\bibfnamefont {N.}~\bibnamefont {Afshordi}},\
  }\bibfield  {title} {\bibinfo {title} {{Echoes from braneworld black
  holes}},\ }\href {https://doi.org/10.1103/PhysRevD.101.104014} {\bibfield
  {journal} {\bibinfo  {journal} {Phys. Rev. D}\ }\textbf {\bibinfo {volume}
  {101}},\ \bibinfo {pages} {104014} (\bibinfo {year} {2020}{\natexlab{b}})},\
  \Eprint {https://arxiv.org/abs/2001.01301} {arXiv:2001.01301 [gr-qc]}
  \BibitemShut {NoStop}%
\bibitem [{\citenamefont {Shiromizu}\ \emph {et~al.}(2000)\citenamefont
  {Shiromizu}, \citenamefont {Maeda},\ and\ \citenamefont
  {Sasaki}}]{Shiromizu:1999wj}%
  \BibitemOpen
  \bibfield  {author} {\bibinfo {author} {\bibfnamefont {T.}~\bibnamefont
  {Shiromizu}}, \bibinfo {author} {\bibfnamefont {K.-i.}\ \bibnamefont
  {Maeda}},\ and\ \bibinfo {author} {\bibfnamefont {M.}~\bibnamefont
  {Sasaki}},\ }\bibfield  {title} {\bibinfo {title} {{The Einstein equation on
  the 3-brane world}},\ }\href {https://doi.org/10.1103/PhysRevD.62.024012}
  {\bibfield  {journal} {\bibinfo  {journal} {Phys. Rev. D}\ }\textbf {\bibinfo
  {volume} {62}},\ \bibinfo {pages} {024012} (\bibinfo {year} {2000})},\
  \Eprint {https://arxiv.org/abs/gr-qc/9910076} {arXiv:gr-qc/9910076}
  \BibitemShut {NoStop}%
\bibitem [{\citenamefont {Aliev}\ and\ \citenamefont
  {Talazan}(2009)}]{Aliev:2009cg}%
  \BibitemOpen
  \bibfield  {author} {\bibinfo {author} {\bibfnamefont {A.~N.}\ \bibnamefont
  {Aliev}}\ and\ \bibinfo {author} {\bibfnamefont {P.}~\bibnamefont
  {Talazan}},\ }\bibfield  {title} {\bibinfo {title} {{Gravitational Effects of
  Rotating Braneworld Black Holes}},\ }\href
  {https://doi.org/10.1103/PhysRevD.80.044023} {\bibfield  {journal} {\bibinfo
  {journal} {Phys. Rev. D}\ }\textbf {\bibinfo {volume} {80}},\ \bibinfo
  {pages} {044023} (\bibinfo {year} {2009})},\ \Eprint
  {https://arxiv.org/abs/0906.1465} {arXiv:0906.1465 [gr-qc]} \BibitemShut
  {NoStop}%
\bibitem [{\citenamefont {Horvath}\ and\ \citenamefont
  {Gergely}(2013)}]{Horvath:2012ru}%
  \BibitemOpen
  \bibfield  {author} {\bibinfo {author} {\bibfnamefont {Z.}~\bibnamefont
  {Horvath}}\ and\ \bibinfo {author} {\bibfnamefont {L.~A.}\ \bibnamefont
  {Gergely}},\ }\bibfield  {title} {\bibinfo {title} {{Black hole tidal charge
  constrained by strong gravitational lensing}},\ }\href
  {https://doi.org/10.1002/asna.201211992} {\bibfield  {journal} {\bibinfo
  {journal} {Astron. Nachr.}\ }\textbf {\bibinfo {volume} {334}},\ \bibinfo
  {pages} {1047} (\bibinfo {year} {2013})},\ \Eprint
  {https://arxiv.org/abs/1203.6576} {arXiv:1203.6576 [gr-qc]} \BibitemShut
  {NoStop}%
\bibitem [{\citenamefont {Zakharov}(2018)}]{Zakharov:2018awx}%
  \BibitemOpen
  \bibfield  {author} {\bibinfo {author} {\bibfnamefont {A.~F.}\ \bibnamefont
  {Zakharov}},\ }\bibfield  {title} {\bibinfo {title} {{Constraints on tidal
  charge of the supermassive black hole at the Galactic Center with
  trajectories of bright stars}},\ }\href
  {https://doi.org/10.1140/epjc/s10052-018-6166-5} {\bibfield  {journal}
  {\bibinfo  {journal} {Eur. Phys. J. C}\ }\textbf {\bibinfo {volume} {78}},\
  \bibinfo {pages} {689} (\bibinfo {year} {2018})},\ \Eprint
  {https://arxiv.org/abs/1804.10374} {arXiv:1804.10374 [gr-qc]} \BibitemShut
  {NoStop}%
\bibitem [{\citenamefont {Banerjee}\ \emph {et~al.}(2019)\citenamefont
  {Banerjee}, \citenamefont {Chakraborty},\ and\ \citenamefont
  {SenGupta}}]{Banerjee:2019sae}%
  \BibitemOpen
  \bibfield  {author} {\bibinfo {author} {\bibfnamefont {I.}~\bibnamefont
  {Banerjee}}, \bibinfo {author} {\bibfnamefont {S.}~\bibnamefont
  {Chakraborty}},\ and\ \bibinfo {author} {\bibfnamefont {S.}~\bibnamefont
  {SenGupta}},\ }\bibfield  {title} {\bibinfo {title} {{Decoding signatures of
  extra dimensions and estimating spin of quasars from the continuum
  spectrum}},\ }\href {https://doi.org/10.1103/PhysRevD.100.044045} {\bibfield
  {journal} {\bibinfo  {journal} {Phys. Rev. D}\ }\textbf {\bibinfo {volume}
  {100}},\ \bibinfo {pages} {044045} (\bibinfo {year} {2019})},\ \Eprint
  {https://arxiv.org/abs/1905.08043} {arXiv:1905.08043 [gr-qc]} \BibitemShut
  {NoStop}%
\bibitem [{\citenamefont {Neves}(2020)}]{Neves:2020doc}%
  \BibitemOpen
  \bibfield  {author} {\bibinfo {author} {\bibfnamefont {J.~C.~S.}\
  \bibnamefont {Neves}},\ }\bibfield  {title} {\bibinfo {title} {{Constraining
  the tidal charge of brane black holes using their shadows}},\ }\href
  {https://doi.org/10.1140/epjc/s10052-020-8321-z} {\bibfield  {journal}
  {\bibinfo  {journal} {Eur. Phys. J. C}\ }\textbf {\bibinfo {volume} {80}},\
  \bibinfo {pages} {717} (\bibinfo {year} {2020})},\ \Eprint
  {https://arxiv.org/abs/2005.00483} {arXiv:2005.00483 [gr-qc]} \BibitemShut
  {NoStop}%
\bibitem [{\citenamefont {Chakraborty}\ \emph {et~al.}(2021)\citenamefont
  {Chakraborty}, \citenamefont {Datta},\ and\ \citenamefont
  {Sau}}]{Chakraborty:2021gdf}%
  \BibitemOpen
  \bibfield  {author} {\bibinfo {author} {\bibfnamefont {S.}~\bibnamefont
  {Chakraborty}}, \bibinfo {author} {\bibfnamefont {S.}~\bibnamefont {Datta}},\
  and\ \bibinfo {author} {\bibfnamefont {S.}~\bibnamefont {Sau}},\ }\bibfield
  {title} {\bibinfo {title} {{Tidal heating of black holes and exotic compact
  objects on the brane}},\ }\href@noop {} {\  (\bibinfo {year} {2021})},\
  \Eprint {https://arxiv.org/abs/2103.12430} {arXiv:2103.12430 [gr-qc]}
  \BibitemShut {NoStop}%
\bibitem [{\citenamefont {Abbott}\ \emph
  {et~al.}(2016{\natexlab{a}})\citenamefont {Abbott} \emph
  {et~al.}}]{Abbott:2016blz}%
  \BibitemOpen
  \bibfield  {author} {\bibinfo {author} {\bibfnamefont {B.~P.}\ \bibnamefont
  {Abbott}} \emph {et~al.} (\bibinfo {collaboration} {LIGO Scientific,
  Virgo}),\ }\bibfield  {title} {\bibinfo {title} {{Observation of
  Gravitational Waves from a Binary Black Hole Merger}},\ }\href
  {https://doi.org/10.1103/PhysRevLett.116.061102} {\bibfield  {journal}
  {\bibinfo  {journal} {Phys. Rev. Lett.}\ }\textbf {\bibinfo {volume} {116}},\
  \bibinfo {pages} {061102} (\bibinfo {year} {2016}{\natexlab{a}})},\ \Eprint
  {https://arxiv.org/abs/1602.03837} {arXiv:1602.03837 [gr-qc]} \BibitemShut
  {NoStop}%
\bibitem [{\citenamefont {Abbott}\ \emph
  {et~al.}(2016{\natexlab{b}})\citenamefont {Abbott} \emph
  {et~al.}}]{TheLIGOScientific:2016src}%
  \BibitemOpen
  \bibfield  {author} {\bibinfo {author} {\bibfnamefont {B.~P.}\ \bibnamefont
  {Abbott}} \emph {et~al.} (\bibinfo {collaboration} {LIGO Scientific,
  Virgo}),\ }\bibfield  {title} {\bibinfo {title} {{Tests of general relativity
  with GW150914}},\ }\href {https://doi.org/10.1103/PhysRevLett.116.221101}
  {\bibfield  {journal} {\bibinfo  {journal} {Phys. Rev. Lett.}\ }\textbf
  {\bibinfo {volume} {116}},\ \bibinfo {pages} {221101} (\bibinfo {year}
  {2016}{\natexlab{b}})},\ \bibinfo {note} {[Erratum: Phys.Rev.Lett. 121,
  129902 (2018)]},\ \Eprint {https://arxiv.org/abs/1602.03841}
  {arXiv:1602.03841 [gr-qc]} \BibitemShut {NoStop}%
\bibitem [{\citenamefont {Banerjee}\ \emph {et~al.}(2021)\citenamefont
  {Banerjee}, \citenamefont {Chakraborty},\ and\ \citenamefont
  {SenGupta}}]{Banerjee:2021aln}%
  \BibitemOpen
  \bibfield  {author} {\bibinfo {author} {\bibfnamefont {I.}~\bibnamefont
  {Banerjee}}, \bibinfo {author} {\bibfnamefont {S.}~\bibnamefont
  {Chakraborty}},\ and\ \bibinfo {author} {\bibfnamefont {S.}~\bibnamefont
  {SenGupta}},\ }\bibfield  {title} {\bibinfo {title} {{Looking for extra
  dimensions in the observed quasi-periodic oscillations of black holes}},\
  }\href@noop {} {\  (\bibinfo {year} {2021})},\ \Eprint
  {https://arxiv.org/abs/2105.06636} {arXiv:2105.06636 [gr-qc]} \BibitemShut
  {NoStop}%
\bibitem [{\citenamefont {Teukolsky}(1973)}]{Teukolsky:1973ha}%
  \BibitemOpen
  \bibfield  {author} {\bibinfo {author} {\bibfnamefont {S.~A.}\ \bibnamefont
  {Teukolsky}},\ }\bibfield  {title} {\bibinfo {title} {{Perturbations of a
  rotating black hole. 1. Fundamental equations for gravitational
  electromagnetic and neutrino field perturbations}},\ }\href
  {https://doi.org/10.1086/152444} {\bibfield  {journal} {\bibinfo  {journal}
  {Astrophys. J.}\ }\textbf {\bibinfo {volume} {185}},\ \bibinfo {pages} {635}
  (\bibinfo {year} {1973})}\BibitemShut {NoStop}%
\bibitem [{\citenamefont {Berti}\ and\ \citenamefont
  {Kokkotas}(2005)}]{Berti:2005eb}%
  \BibitemOpen
  \bibfield  {author} {\bibinfo {author} {\bibfnamefont {E.}~\bibnamefont
  {Berti}}\ and\ \bibinfo {author} {\bibfnamefont {K.~D.}\ \bibnamefont
  {Kokkotas}},\ }\bibfield  {title} {\bibinfo {title} {{Quasinormal modes of
  Kerr-Newman black holes: Coupling of electromagnetic and gravitational
  perturbations}},\ }\href {https://doi.org/10.1103/PhysRevD.71.124008}
  {\bibfield  {journal} {\bibinfo  {journal} {Phys. Rev. D}\ }\textbf {\bibinfo
  {volume} {71}},\ \bibinfo {pages} {124008} (\bibinfo {year} {2005})},\
  \Eprint {https://arxiv.org/abs/gr-qc/0502065} {arXiv:gr-qc/0502065}
  \BibitemShut {NoStop}%
\bibitem [{\citenamefont {Dudley}\ and\ \citenamefont
  {Finley}(1979)}]{Dudley:1978vd}%
  \BibitemOpen
  \bibfield  {author} {\bibinfo {author} {\bibfnamefont {A.~L.}\ \bibnamefont
  {Dudley}}\ and\ \bibinfo {author} {\bibfnamefont {J.~D.}\ \bibnamefont
  {Finley}, \bibfnamefont {III}},\ }\bibfield  {title} {\bibinfo {title}
  {{Covariant Perturbed Wave Equations in Arbitrary Type $D$ Backgrounds}},\
  }\href {https://doi.org/10.1063/1.524064} {\bibfield  {journal} {\bibinfo
  {journal} {J. Math. Phys.}\ }\textbf {\bibinfo {volume} {20}},\ \bibinfo
  {pages} {311} (\bibinfo {year} {1979})}\BibitemShut {NoStop}%
\bibitem [{\citenamefont {Dudley}\ and\ \citenamefont
  {Finley}(1977)}]{PhysRevLett.38.1505}%
  \BibitemOpen
  \bibfield  {author} {\bibinfo {author} {\bibfnamefont {A.~L.}\ \bibnamefont
  {Dudley}}\ and\ \bibinfo {author} {\bibfnamefont {J.~D.}\ \bibnamefont
  {Finley}},\ }\bibfield  {title} {\bibinfo {title} {Separation of wave
  equations for perturbations of general type-$d$ space-times},\ }\href
  {https://doi.org/10.1103/PhysRevLett.38.1505} {\bibfield  {journal} {\bibinfo
   {journal} {Phys. Rev. Lett.}\ }\textbf {\bibinfo {volume} {38}},\ \bibinfo
  {pages} {1505} (\bibinfo {year} {1977})}\BibitemShut {NoStop}%
\bibitem [{\citenamefont {Mark}\ \emph {et~al.}(2015)\citenamefont {Mark},
  \citenamefont {Yang}, \citenamefont {Zimmerman},\ and\ \citenamefont
  {Chen}}]{Mark:2014aja}%
  \BibitemOpen
  \bibfield  {author} {\bibinfo {author} {\bibfnamefont {Z.}~\bibnamefont
  {Mark}}, \bibinfo {author} {\bibfnamefont {H.}~\bibnamefont {Yang}}, \bibinfo
  {author} {\bibfnamefont {A.}~\bibnamefont {Zimmerman}},\ and\ \bibinfo
  {author} {\bibfnamefont {Y.}~\bibnamefont {Chen}},\ }\bibfield  {title}
  {\bibinfo {title} {{Quasinormal modes of weakly charged Kerr-Newman
  spacetimes}},\ }\href {https://doi.org/10.1103/PhysRevD.91.044025} {\bibfield
   {journal} {\bibinfo  {journal} {Phys. Rev. D}\ }\textbf {\bibinfo {volume}
  {91}},\ \bibinfo {pages} {044025} (\bibinfo {year} {2015})},\ \Eprint
  {https://arxiv.org/abs/1409.5800} {arXiv:1409.5800 [gr-qc]} \BibitemShut
  {NoStop}%
\bibitem [{\citenamefont {Dias}\ \emph {et~al.}(2015)\citenamefont {Dias},
  \citenamefont {Godazgar},\ and\ \citenamefont {Santos}}]{Dias:2015wqa}%
  \BibitemOpen
  \bibfield  {author} {\bibinfo {author} {\bibfnamefont {O.~J.~C.}\
  \bibnamefont {Dias}}, \bibinfo {author} {\bibfnamefont {M.}~\bibnamefont
  {Godazgar}},\ and\ \bibinfo {author} {\bibfnamefont {J.~E.}\ \bibnamefont
  {Santos}},\ }\bibfield  {title} {\bibinfo {title} {{Linear Mode Stability of
  the Kerr-Newman Black Hole and Its Quasinormal Modes}},\ }\href
  {https://doi.org/10.1103/PhysRevLett.114.151101} {\bibfield  {journal}
  {\bibinfo  {journal} {Phys. Rev. Lett.}\ }\textbf {\bibinfo {volume} {114}},\
  \bibinfo {pages} {151101} (\bibinfo {year} {2015})},\ \Eprint
  {https://arxiv.org/abs/1501.04625} {arXiv:1501.04625 [gr-qc]} \BibitemShut
  {NoStop}%
\bibitem [{\citenamefont {Kanno}\ and\ \citenamefont
  {Soda}(2004)}]{Kanno:2003au}%
  \BibitemOpen
  \bibfield  {author} {\bibinfo {author} {\bibfnamefont {S.}~\bibnamefont
  {Kanno}}\ and\ \bibinfo {author} {\bibfnamefont {J.}~\bibnamefont {Soda}},\
  }\bibfield  {title} {\bibinfo {title} {{Rotating black string and effective
  Teukolsky equation in brane world}},\ }\href
  {https://doi.org/10.1088/0264-9381/21/7/012} {\bibfield  {journal} {\bibinfo
  {journal} {Class. Quant. Grav.}\ }\textbf {\bibinfo {volume} {21}},\ \bibinfo
  {pages} {1915} (\bibinfo {year} {2004})},\ \Eprint
  {https://arxiv.org/abs/gr-qc/0311074} {arXiv:gr-qc/0311074} \BibitemShut
  {NoStop}%
\bibitem [{\citenamefont {Kanno}\ and\ \citenamefont
  {Soda}(2003)}]{Kanno:2003sc}%
  \BibitemOpen
  \bibfield  {author} {\bibinfo {author} {\bibfnamefont {S.}~\bibnamefont
  {Kanno}}\ and\ \bibinfo {author} {\bibfnamefont {J.}~\bibnamefont {Soda}},\
  }\bibfield  {title} {\bibinfo {title} {{Effective Teukolsky equation on the
  brane}},\ }in\ \href@noop {} {\emph {\bibinfo {booktitle} {{6th RESCEU
  International Symposium on Frontier in Astroparticle Physics and
  Cosmology}}}}\ (\bibinfo {year} {2003})\ \Eprint
  {https://arxiv.org/abs/gr-qc/0312098} {arXiv:gr-qc/0312098} \BibitemShut
  {NoStop}%
\bibitem [{\citenamefont {Leaver}(1985)}]{Leaver:1985ax}%
  \BibitemOpen
  \bibfield  {author} {\bibinfo {author} {\bibfnamefont {E.~W.}\ \bibnamefont
  {Leaver}},\ }\bibfield  {title} {\bibinfo {title} {{An Analytic
  representation for the quasi normal modes of Kerr black holes}},\ }\href
  {https://doi.org/10.1098/rspa.1985.0119} {\bibfield  {journal} {\bibinfo
  {journal} {Proc. Roy. Soc. Lond. A}\ }\textbf {\bibinfo {volume} {402}},\
  \bibinfo {pages} {285} (\bibinfo {year} {1985})}\BibitemShut {NoStop}%
\bibitem [{\citenamefont {Seahra}\ \emph {et~al.}(2005)\citenamefont {Seahra},
  \citenamefont {Clarkson},\ and\ \citenamefont {Maartens}}]{Seahra:2004fg}%
  \BibitemOpen
  \bibfield  {author} {\bibinfo {author} {\bibfnamefont {S.~S.}\ \bibnamefont
  {Seahra}}, \bibinfo {author} {\bibfnamefont {C.}~\bibnamefont {Clarkson}},\
  and\ \bibinfo {author} {\bibfnamefont {R.}~\bibnamefont {Maartens}},\
  }\bibfield  {title} {\bibinfo {title} {{Detecting extra dimensions with
  gravity wave spectroscopy: the black string brane-world}},\ }\href
  {https://doi.org/10.1103/PhysRevLett.94.121302} {\bibfield  {journal}
  {\bibinfo  {journal} {Phys. Rev. Lett.}\ }\textbf {\bibinfo {volume} {94}},\
  \bibinfo {pages} {121302} (\bibinfo {year} {2005})},\ \Eprint
  {https://arxiv.org/abs/gr-qc/0408032} {arXiv:gr-qc/0408032} \BibitemShut
  {NoStop}%
\bibitem [{\citenamefont {Andriot}\ and\ \citenamefont
  {Lucena~G\'omez}(2017)}]{Andriot:2017oaz}%
  \BibitemOpen
  \bibfield  {author} {\bibinfo {author} {\bibfnamefont {D.}~\bibnamefont
  {Andriot}}\ and\ \bibinfo {author} {\bibfnamefont {G.}~\bibnamefont
  {Lucena~G\'omez}},\ }\bibfield  {title} {\bibinfo {title} {{Signatures of
  extra dimensions in gravitational waves}},\ }\href
  {https://doi.org/10.1088/1475-7516/2017/06/048} {\bibfield  {journal}
  {\bibinfo  {journal} {JCAP}\ }\textbf {\bibinfo {volume} {06}},\ \bibinfo
  {pages} {048}},\ \bibinfo {note} {[Erratum: JCAP 05, E01 (2019)]},\ \Eprint
  {https://arxiv.org/abs/1704.07392} {arXiv:1704.07392 [hep-th]} \BibitemShut
  {NoStop}%
\bibitem [{\citenamefont {Collaboration}\ and\ \citenamefont
  {Collaboration}(2020)}]{public_data_release_GWTC2_TGR}%
  \BibitemOpen
  \bibfield  {author} {\bibinfo {author} {\bibfnamefont {L.~S.}\ \bibnamefont
  {Collaboration}}\ and\ \bibinfo {author} {\bibfnamefont {V.}~\bibnamefont
  {Collaboration}},\ }\href {https://doi.org/10.7935/903S-GX73} {\bibinfo
  {title} {Data release for tests of general relativity with gwtc-2}} (\bibinfo
  {year} {2020})\BibitemShut {NoStop}%
\bibitem [{\citenamefont {Carullo}\ \emph {et~al.}(2019)\citenamefont
  {Carullo}, \citenamefont {Del~Pozzo},\ and\ \citenamefont
  {Veitch}}]{Carullo:2019flw}%
  \BibitemOpen
  \bibfield  {author} {\bibinfo {author} {\bibfnamefont {G.}~\bibnamefont
  {Carullo}}, \bibinfo {author} {\bibfnamefont {W.}~\bibnamefont {Del~Pozzo}},\
  and\ \bibinfo {author} {\bibfnamefont {J.}~\bibnamefont {Veitch}},\
  }\bibfield  {title} {\bibinfo {title} {{Observational Black Hole
  Spectroscopy: A time-domain multimode analysis of GW150914}},\ }\href
  {https://doi.org/10.1103/PhysRevD.99.123029} {\bibfield  {journal} {\bibinfo
  {journal} {Phys. Rev. D}\ }\textbf {\bibinfo {volume} {99}},\ \bibinfo
  {pages} {123029} (\bibinfo {year} {2019})},\ \bibinfo {note} {[Erratum:
  Phys.Rev.D 100, 089903 (2019)]},\ \Eprint {https://arxiv.org/abs/1902.07527}
  {arXiv:1902.07527 [gr-qc]} \BibitemShut {NoStop}%
\bibitem [{\citenamefont {Isi}\ \emph {et~al.}(2019)\citenamefont {Isi},
  \citenamefont {Giesler}, \citenamefont {Farr}, \citenamefont {Scheel},\ and\
  \citenamefont {Teukolsky}}]{Isi:2019aib}%
  \BibitemOpen
  \bibfield  {author} {\bibinfo {author} {\bibfnamefont {M.}~\bibnamefont
  {Isi}}, \bibinfo {author} {\bibfnamefont {M.}~\bibnamefont {Giesler}},
  \bibinfo {author} {\bibfnamefont {W.~M.}\ \bibnamefont {Farr}}, \bibinfo
  {author} {\bibfnamefont {M.~A.}\ \bibnamefont {Scheel}},\ and\ \bibinfo
  {author} {\bibfnamefont {S.~A.}\ \bibnamefont {Teukolsky}},\ }\bibfield
  {title} {\bibinfo {title} {{Testing the no-hair theorem with GW150914}},\
  }\href {https://doi.org/10.1103/PhysRevLett.123.111102} {\bibfield  {journal}
  {\bibinfo  {journal} {Phys. Rev. Lett.}\ }\textbf {\bibinfo {volume} {123}},\
  \bibinfo {pages} {111102} (\bibinfo {year} {2019})},\ \Eprint
  {https://arxiv.org/abs/1905.00869} {arXiv:1905.00869 [gr-qc]} \BibitemShut
  {NoStop}%
\bibitem [{\citenamefont {London}\ \emph {et~al.}(2014)\citenamefont {London},
  \citenamefont {Shoemaker},\ and\ \citenamefont {Healy}}]{London:2014cma}%
  \BibitemOpen
  \bibfield  {author} {\bibinfo {author} {\bibfnamefont {L.}~\bibnamefont
  {London}}, \bibinfo {author} {\bibfnamefont {D.}~\bibnamefont {Shoemaker}},\
  and\ \bibinfo {author} {\bibfnamefont {J.}~\bibnamefont {Healy}},\ }\bibfield
   {title} {\bibinfo {title} {{Modeling ringdown: Beyond the fundamental
  quasinormal modes}},\ }\href {https://doi.org/10.1103/PhysRevD.90.124032}
  {\bibfield  {journal} {\bibinfo  {journal} {Phys. Rev. D}\ }\textbf {\bibinfo
  {volume} {90}},\ \bibinfo {pages} {124032} (\bibinfo {year} {2014})},\
  \bibinfo {note} {[Erratum: Phys.Rev.D 94, 069902 (2016)]},\ \Eprint
  {https://arxiv.org/abs/1404.3197} {arXiv:1404.3197 [gr-qc]} \BibitemShut
  {NoStop}%
\bibitem [{\citenamefont {London}(2020)}]{London:2018gaq}%
  \BibitemOpen
  \bibfield  {author} {\bibinfo {author} {\bibfnamefont {L.~T.}\ \bibnamefont
  {London}},\ }\bibfield  {title} {\bibinfo {title} {{Modeling ringdown. II.
  Aligned-spin binary black holes, implications for data analysis and
  fundamental theory}},\ }\href {https://doi.org/10.1103/PhysRevD.102.084052}
  {\bibfield  {journal} {\bibinfo  {journal} {Phys. Rev. D}\ }\textbf {\bibinfo
  {volume} {102}},\ \bibinfo {pages} {084052} (\bibinfo {year} {2020})},\
  \Eprint {https://arxiv.org/abs/1801.08208} {arXiv:1801.08208 [gr-qc]}
  \BibitemShut {NoStop}%
\bibitem [{\citenamefont {Ghosh}\ \emph {et~al.}(2021)\citenamefont {Ghosh},
  \citenamefont {Brito},\ and\ \citenamefont {Buonanno}}]{Ghosh:2021mrv}%
  \BibitemOpen
  \bibfield  {author} {\bibinfo {author} {\bibfnamefont {A.}~\bibnamefont
  {Ghosh}}, \bibinfo {author} {\bibfnamefont {R.}~\bibnamefont {Brito}},\ and\
  \bibinfo {author} {\bibfnamefont {A.}~\bibnamefont {Buonanno}},\ }\bibfield
  {title} {\bibinfo {title} {{Constraints on quasi-normal-mode frequencies with
  LIGO-Virgo binary-black-hole observations}},\ }\href@noop {} {\  (\bibinfo
  {year} {2021})},\ \Eprint {https://arxiv.org/abs/2104.01906}
  {arXiv:2104.01906 [gr-qc]} \BibitemShut {NoStop}%
\bibitem [{\citenamefont {Carullo}(2021)}]{Carullo:2021dui}%
  \BibitemOpen
  \bibfield  {author} {\bibinfo {author} {\bibfnamefont {G.}~\bibnamefont
  {Carullo}},\ }\bibfield  {title} {\bibinfo {title} {{Accelerating modified
  gravity detection from gravitational-wave observations using the Parametrized
  ringdown spin expansion coefficients formalism}},\ }\href@noop {} {\
  (\bibinfo {year} {2021})},\ \Eprint {https://arxiv.org/abs/2102.05939}
  {arXiv:2102.05939 [gr-qc]} \BibitemShut {NoStop}%
\end{thebibliography}%

\bibliographystyle{./utphys1}
\end{document}